\documentclass[prb,superscriptaddress,twocolumn,preprintnumbers,a4,showpacs,floatfix]{revtex4}

\usepackage{psfrag, tabls, dcolumn}
\usepackage[sort&compress]{natbib}
\usepackage[pdftex]{color, graphicx}

\usepackage{mathtools}

\usepackage{multirow}
\usepackage{booktabs}
\usepackage{tabularx}
\usepackage{longtable}
\usepackage{supertabular}

\usepackage{bm, amsmath, amsfonts, amsbsy}

\newcolumntype{C}{>{$\displaystyle}c<{$}}
\newcolumntype{L}{>{$\displaystyle}l<{$}}
\newcolumntype{R}{>{$\displaystyle}r<{$}}

\newcommand{\I}{i}
\renewcommand{\exp}[1]{\operatorname{e}^{#1}}
\newcommand{\B}{\textcolor{blue}}

\def\unity{\R{\mathbb{I}}}      
\def\unity{{\mathbf{1}}}        

\newcommand{\ssref}[1]{Subsection~\ref{SS:#1}}

\newcommand{\FIG}[1]{Figure~\ref{#1}}

\newcommand{\tab}[1]{(Table~\ref{#1})}
\newcommand{\TAB}[1]{Table~\ref{#1}}

\def\eg{\textit{e.g.} }
\def\ie{\textit{i.e.} }
\def\ieo{\textit{i.e.}}
\def\cf{\textit{cf.} }

\def\wf{wave function }

\def\wfs{wave functions }

\def\wrt{with respect to }

\def\rbloch{revised Bloch's theorem }
\def\rblocho{revised Bloch's theorem}

\def\Rblocho{Revised Bloch's theorem}

\newcommand{\vTab}{{\vphantom{X^{X^{X^{X}}}_{X_{X}}}}}

\newcommand\TrX{\operatorname{Tr}}
\newcommand\Tr[1]{\TrX\left[#1\right]}
\newcommand\Trp[2]{\TrX_{#1}\left[#2\right]}

\def\Ncells{\mathcal{N}}                

\newcommand{\Edef}{\equiv}      

\def\aln{\al_{\nb}}                                 

\def\ombn{\omb_{\nb}}                               
\def\omn{\om_{\nb}}                                 
\def\Tn{\Tb_{\nb}}                                  

\def\RA{\Rb_A}                                          
\def\RI{\Rb_I}                                          
\def\RJ{\Rb_J}                                          

\def\Rmu{\Rb_{\mu}}                                 

\def\Rmu{\Rb_{\mu}}                                 %
\def\RnA{\Rb^{\nb}_A}                               
\def\RnI{\Rb^{\nb}_I}

\def\RnJ{\Rb^{\nb}_J}
\def\RmJ{\Rb^{\mb}_J}

\def\RXJ#1{\Rb^{#1}_J}

\def\Rnmup{\Rb^{\nb}_{\mu'}}            
\def\Rnnu{\Rb^{\nb}_{\nu}}          
\def\Rmup{\Rb_{\mu'}}           

\def\RNnIJ{R^{\phantom{\nb}\nb}_{IJ}}                   
\def\RNnJI{R^{\phantom{\nb}\nb}_{JI}}                   
\def\RNXIJ#1{R^{\phantom{\nb}#1}_{IJ}}              
\def\RNnII{R^{\phantom{\nb}\nb}_{II}}                   

\def\RNIJ{R_{IJ}}

\def\RnIJ{\Rb^{\phantom{\nb}\nb}_{IJ}}
\def\RXIJ#1{\Rb^{\phantom{\nb}#1}_{IJ}}

\def\RhnIJ{\Rhb^{\phantom{\nb}\nb}_{IJ}}

\def\RhnII{\Rhb^{\phantom{\nb}\nb}_{II  }}                  
\def\RhnJI{\Rhb^{\phantom{\nb}\nb}_{JI}}                    

\def\RhXIJ#1{\Rhb^{\phantom{\nb}#1}_{IJ}}

\def\Rnnu{\Rb^{\nb}_{\nu}}          %

\newcommand{\matr}{\left( \begin{array}}
\newcommand{\ematr}{\end{array} \right)}

\def\HM{\mathbf{H}}                 
\def\HpM{\mathbf{H'}}

\def\rhoM{\bm \rho}

\def\oOrn{\Oover{1}{\sqrt{\Ncells}}}
\def\oOn{\Oover{1}{{\Ncells}}}

\def\ChemOhf{
    \{
        \Ket{\pfi^{0}_{\mu}}
    \}
    =
    \{
        &
        \Ket{s}, \,
        \Ket{p_x},
        \Ket{p_y},
        \Ket{p_z},
    \nl &
        \Ket{d_{x^2-y^2}},
        \Ket{d_{xy}},
        \Ket{d_{yz}},
        \Ket{d_{zx}},
        \Ket{d_{z^2}}, \,
        \dots
    \}
}

\newcommand{\fii}[2]{\pfi^{#1}_{#2}}

\def\Caln{\cos(\aln)}
\def\Saln{\sin(\aln)}
\def\CTaln{\cos(2\aln)}     
\def\STaln{\sin(2\aln)}     

\def\bit{\begin{itemize}}
\def\eit{\end{itemize}}
\def\ben#1{\begin{enumerate}[#1]}
\def\een{\end{enumerate}}

\def\bcols{\begin{columns}}
\def\ecols{\end{columns}}
\def\bcol{\begin{column}}
\def\ecol{\end{column}}
\def\colb#1{\begin{columns}\begin{column}{#1}}

\def\cole{\end{column}\end{columns}}

\newcommand{\be}{\begin{equation}}
\newcommand{\ee}{\end{equation}}
\newcommand{\bea}{\begin{equation*}}
\newcommand{\eea}{\end{equation*}}
\newcommand{\ba}{\begin{array}}
\newcommand{\ea}{\end{array}}
\newcommand{\beqa}{\begin{eqnarray}}
\newcommand{\eeqa}{\end{eqnarray}}
\newcommand{\beqaa}{\begin{eqnarray*}}
\newcommand{\eeqaa}{\end{eqnarray*}}

\newcommand{\lbl}[1]{\label{e:#1}}
\newcommand{\eref}[1]{\eqref{e:#1}}                                 
\newcommand{\er}[1]{~\eqref{e:#1}}                                  
\newcommand{\eq}[1]{Eq.~\eref{#1}}

\newcommand{\baaa}{\begin{align}}
\newcommand{\eaaa}{\end{align}}

\def\nl{\nonumber\\}        
\def\nn{\nonumber}        

\newcommand{\al}{\alpha}
\newcommand{\om}{\omega}
\newcommand{\Be}{\beta}
\newcommand{\ka}{\kappa}

\newcommand{\pfi}{\varphi}
\newcommand{\eps}{\varepsilon}

\renewcommand{\bm}[1]{\boldsymbol{\mathbf{#1}}}

\newcommand{\nab}{\bm \nabla}

\newcommand{\bb}{\bm b}

\newcommand{\db}{\bm d}

\newcommand{\kb}{\bm k}

\newcommand{\rb}{\bm r}

\newcommand{\fb}{\bm f}

\newcommand{\nb}{\bm n}
\newcommand{\mb}{\bm m}

\newcommand{\Fb}{\bm F}

\newcommand{\Mb}{\bm M}
\newcommand{\Lb}{\bm L}

\newcommand{\Rb}{\bm R}
\newcommand{\Sb}{\bm S}
\newcommand{\Hb}{\bm H}
\newcommand{\Tb}{\bm T}

\def\kab{\bm \ka}
\def\alb{\bm \al}
\def\omb{\bm \omega}
\def\hib{\bm \chi}
\def\rhob{\bm \rho}

\newcommand{\dert}{{\rm d^3}}

\newcommand{\dtr}{{\dert r}}

\def\Itr{\int\!\dtr\,}

\newcommand{\half}{{\Oover{1}{2}}}

\def\Oover#1#2{\hbox{$\textstyle {#1\over #2}$}}

\def\Lh{\hat L}
\def\ph{\hat p}

\def\Hh{\hat H}

\def\kah{\hat \ka}

\def\eh{\hat e}

\def\phb{\bm \ph}

\def\kahb{\bm \kah}
\def\ehb{\bm \eh}

\def\zbh{\bm{\hat z}}
\def\Rhb{\hat{\Rb}} 

\def\inv{^{-1}}

\def\SymNot#1{\mathcal{#1}} 

\def\Sym{\SymNot S}     
\def\SymIn{\Sym^{-n}}   
\def\SymInb{\Sym^{-\nb}}    
\def\Symn{\Sym^n}
\def\Symnb{\Sym^{\nb}}
\def\Symmb{\Sym^{\mb}}

\def\Symmb{\Sym^{\mb}}          
\def\SymX#1{\Sym^{#1}}
\def\SymY#1#2{\Sym^{#1}_{#2}}

\def\Tra#1{\SymNot T^{#1}}

\def\Rot{\Roti}                                         

\def\Roti#1{\SymNot R (\bm{#1})\,}          

\def\RotX#1{\SymNot R (#1)\,}           

\def\DSymm{\mathnormal{\hat{D}}}
\def\DdSymm{\DSymm^{\dagger}}       

\def\DSym{\DSymm(\Sym)}
\def\DSymn{\DSymm(\Symn)}
\def\DSymnb{\DSymm(\Symnb)}
\def\DISymnb{\DSymm\inv(\Symnb)}        
\def\DdSymnb{\DdSymm(\Symnb)}       
\def\DSymnbI{\DSymm(\SymInb)}       
\def\DSymX#1{\DSymm(\SymX{#1})}
\def\DSymY#1#2{\DSymm(\SymY{#1}{#2})}

\def\DTra#1{\DSymm(\Tra{#1})}

\def\SymInd#1{\vphantom{#1}}                        

\def\DTTra#1{\DSymm\SymInd{t}(\Tra{#1})}
\def\DTdTra#1{\DdSymm\SymInd{t}(\Tra{#1})}
\def\DRRot#1{\DSymm\SymInd{r}(\Rot{#1})}
\def\DSSymnb{\DSymm\SymInd{s}(\Symnb)}

\def\DRRotM#1#2{D^0_{#2}(\Rot{#1})\,}

\def\DRotMX#1{\mathbf{D}_{0}^{(#1)}}                

\def\DRotMM#1{\mathbf{D}^{0}\SymInd{r}(\Rot{#1})}               

\def\Bra#1{\langle#1|}                          
\def\Ket#1{|#1\rangle}                          
\def\BK#1#2{\langle#1|#2\rangle}        

\def\Kpsi{\Ket{\psi}}                           
\def\BKpsi#1{\BK{#1}{\psi}}                 

\def\Brb{\Bra{\rb}}                                 

\def\BfiiX#1#2{\Bra{\fii{#1}{#2}}}
\def\KfiiX#1#2{\Ket{\fii{#1}{#2}}}

\def\kahDn{\kahb\cdot\nb}               
\def\kaDn{\kab\cdot\nb}               
\def\kaDm{\kab\cdot\mb}               

\newcommand\smallurl[1]{{\tiny \url{#1}}}

\usepackage{mdwlist}
\usepackage{units}
\usepackage{color}
\usepackage[normalem]{ulem}

\begin{document}

\title{Revised Periodic Boundary Conditions: \\
Fundamentals, Electrostatics,
and the Tight-Binding Approximation}

\author{Oleg O. Kit}
\affiliation{NanoScience Center, Department of Physics, University of Jyv\"askyl\"a, 40014 Jyv\"askyl\"a, Finland}
\author{Lars Pastewka}
\altaffiliation{{Present address: Department of Physics and Astronomy, Johns Hopkins
  University, Baltimore, MD 21218, USA}}
\affiliation{Fraunhofer Institute for Mechanics of Materials IWM, W\"ohlerstra\ss e 11, 79108 Freiburg, Germany}
\author{Pekka Koskinen}
\email[Corresponding author, email: ]{pekka.koskinen@iki.fi}
\affiliation{NanoScience Center, Department of Physics, University of Jyv\"askyl\"a, 40014 Jyv\"askyl\"a, Finland}


\begin{abstract}
Many nanostructures today are low-dimensional and flimsy, and therefore get easily distorted. Distortion-induced symmetry-breaking makes conventional, translation-periodic simulations invalid, which has triggered developments for new methods. Revised periodic boundary conditions (RPBC) is a simple method that enables simulations of complex material distortions, either classically or quantum-mechanically. The mathematical details of this easy-to-implement approach, however, have not been discussed before. Therefore, in this paper we summarize the underlying theory, present the practical details of RPBC, especially related to a non-orthogonal tight-binding formulation, discuss selected features, electrostatics in particular, and
suggest some examples of usage.
We hope this article to give more insight to RPBC, to help and inspire new software implementations capable of exploring the physics and chemistry of distorted nanomaterials.
\end{abstract}

\maketitle

\section{Material symmetries beyond translations}

Translational symmetry and Bloch's theorem has been the backbone of materials research for a long time~\cite{Bloch}. Bloch's theorem was originally associated with simulations of bulk crystals and translational symmetry in three dimensions, and this association is still strong. Together with periodic boundary conditions (PBC), or Born-von~K\'arm\'an boundary conditions, Bloch's theorem has enabled simulating the infinite bulk using a single, minimal unit cell.

Nanoscience, however, has introduced novel material structures that often lack the translational symmetry. These structures include nanotubes, nanowires, ribbons, nanopeapods, DNA, polymers, proteins, to mention only a few examples. Some structures, such as nanotubes, are often modeled with translational symmetry, which is, however, dangerous because of their low-dimensional character and flimsiness---in experiments the real structures get distorted and Bloch's theorem and conventional PBC becomes invalid. To remedy this problem, during the course of time different research groups have independently developed new methodological improvements.

The seminal ideas to use generalized symmetries were presented by White,
Robertson and Mintmire, who used helical (roto-translational) symmetry to simulate model carbon nanotubes (CNTs) using a two-atom unit cell~\cite{CTWhiteDHRobertsonJWMintmire-PhysRevB.47.5485}, which enabled calculating CNT properties that were inaccessible before~\cite{lawler_PRB_06,white_JPCB_05}. A similar approach, also applied to CNTs, was followed by Popov~\cite{popov_NJP_04,popov_PRB_04}. Liu and Ding used rotational symmetry to simulate the electronic structure of carbon nanotori within a tight-binding model, although with a static geometry~\cite{CPLiuJWDing2006}. Dumitric\u a and James presented their clever objective molecular dynamics (OMD), using classical formulation~\cite{ObjMD-TDumitricaRDJames2007}; later, OMD was extended to tight-binding language~\cite{TBintoObjMD-DBZhangMHuaTDumitrica}. Also Cai \emph{et al.} presented boundary conditions for twisting and bending, using nanowires as their target application~\cite{cai_JMPS_08}. These approaches towards generalized symmetries have been applied to various materials and systems, such as Si nanowires~\cite{TBintoObjMD-DBZhangMHuaTDumitrica}, MoS$_2$-nanotubes~\cite{PhysRevLett.104.065502}, nanoribbons\cite{zhang_small_11}, the bending of single-walled~\cite{CPLiuJWDing2006,PhysRevB.80.115418,zhang_AN_10} and multi-walled~\cite{nikiforov_APL_10} CNTs, and vibrational properties of single-walled CNTs~\cite{malola_PRB_08b}.

Recently, in Ref.~\onlinecite{koskinen_PRL_10}, we introduced revised periodic
boundary conditions (RPBC), a unified method to simulate materials with
versatile distortions. This method has a particularly simple formulation, with
both classical and fully quantum-mechanical treatments. It can be used in
conjunction with molecular dynamics and Monte Carlo simulation schemes, it is designed for general distortions and all material systems, and it can be regarded as a generalization of previous work~\cite{CTWhiteDHRobertsonJWMintmire-PhysRevB.47.5485,lawler_PRB_06,white_JPCB_05,popov_NJP_04,popov_PRB_04,CPLiuJWDing2006,ObjMD-TDumitricaRDJames2007,TBintoObjMD-DBZhangMHuaTDumitrica,cai_JMPS_08}. Because
we only presented the overall outline of RPBCs in
Ref.~\onlinecite{koskinen_PRL_10}, the purpose of this paper is, above all, to
be the mathematical and technical companion of that work. We give in-depth formulation of the
approach, suggest few examples of usage, and discuss the long-range
electrostatic interactions carefully, as to provide a
detailed-enough overview to aid implementing and applying RPBC in practice. Examples
of earlier simulations with RPBC include bending of single-walled
CNTs~\cite{PhysRevB.82.193409}, twisting of graphene nanoribbons~\cite{koskinen_PRL_10,koskinen_APL_11}, and spherical wrapping of
graphene~\cite{PhysRevB.82.235420}.

The idea of using general symmetries in material simulations should not be
surprising, as in physics and chemistry symmetry has always played a special role. What is surprising, though, is that symmetries beyond the
translational ones have not quite reached the mainstream of materials
modeling. Using generalized symmetries does not imply material distortions;
they can be used also merely to reduce computational costs. For example, RPBC is
based on revised Bloch's theorem, which in turn is based on the validity of
Bloch's theorem for any cyclic group---an age-old group-theoretical
common knowledge~\cite{Tinkham}. However, it is not until nanoscience and its
low-dimensional distorted structures that have brought the motivation to
finally go beyond the conventional translational symmetry, especially in practical
simulations. We hope this paper could demonstrate how simple the RPBC
formulation is, and what kind of ingredients are required in the
implementation. Using symmetries beyond translation should be in reach also for
the simulation community mainstream.

\section{Revised Bloch's Theorem}

Let us first, to have a common starting point, repeat the translational Bloch's theorem, as it is found in all solid state textbooks~\cite{Marder}. In its conventional form, Bloch's theorem
\be
    \psi_{a\kb} (\rb - \Tn) = \exp{-\I\kb\cdot\Tn} \psi_{a\kb} (\rb)
    \lbl{bloch_simple}
\ee
is valid for a system of electrons in a periodic potential $V(\rb) = V(\rb + \Tn)$,
where $\Tn = n_1 \Tb_1 + n_2 \Tb_2 + n_3 \Tb_3$ and $\Tb_i$ are lattice vectors. In Eq.~\eref{bloch_simple}, the \wf $\psi_{a\kb}(\rb)$, labeled with band index $a$ and $\kb$-vector, is the solution to Schr\"odinger equation,
\be
  \Hh \psi_{a\kb}(\rb)
      = \left[\phb^2/2m + V(\rb) \right]\psi_{a\kb}(\rb)
      = \varepsilon \psi_{a\kb}(\rb),
   \lbl{shrodinger}
\ee
where $V(\rb)$ could also be the Kohn-Sham potential of the density functional theory. The system is subjected to periodic boundary conditions
(or Born-von~K\'arm\'an boundary conditions), which mean that translation  by $\Lb_j = M_j\Tb_j$, the system's length in direction $j$, leaves the system unchanged. What has made the theorem \eref{bloch_simple} so powerful, is that the knowledge of the \wf within a single unit cell is sufficient to solve the electronic structure and structural properties of the crystal as a whole.

Next we restate Bloch's theorem~\eref{bloch_simple} in a
form easier to generalize. First, we introduce the notation $\Tra{\nb}\rb \equiv \rb + \Tn$ for the coordinate transformation of translation, with the inverse transformation $\Tra{-\nb}\rb = \rb + \Tb_{-\nb} = \rb - \Tb_{\nb}$. Hence the left-hand side of Eq.~\eref{bloch_simple} becomes $\psi_{a\kb}
(\Tra{-\nb}\rb)$.
Second, we define an operator $\DTra{\nb}$ acting on wave functions that induces the
real-space translation $\Tra{\nb}$.
For this, recall that shifting a function ``forward'' [action by $\DTra{\nb}$] is equivalent to shifting the coordinate system ``backward'' ($\rb' = \Tra{-\nb}\rb$),
\be
  \DTra{\nb}\psi(\rb) = \psi(\Tra{-\nb}\rb).
  \lbl{D_def}
\ee
The way to arrive at \eref{D_def} is the following.
Let us make a forward translation $\Tra{\nb}$ from
the state $\psi$ in the coordinate system $\rb$ to a state $\psi'$ in
the coordinate system $\rb'=\Tra{\nb}\rb$.
The \wf does not change, meaning $\psi(\rb) = \psi'(\rb')$ or $ \psi(\rb) = \psi'(\Tra{\nb}\rb)$, which holds for any $\rb$. In particular, it holds for $\rb=\Tra{-\nb}\rb''$, so (dropping the double prime) $\psi(\Tra{-\nb}\rb) = \psi'(\rb)$. Now, since the state $\psi'$ is the state induced by the forward translation, we define $\hat{D}(\Tra{\nb})$ by $\DSymm(\Tra{\nb})\psi(\rb) = \psi'(\rb)$, which leads to \eref{D_def}. With this notation the Bloch's theorem becomes
\be
	\DTra{\nb}\psi_{a\kb}(\rb) = \psi_{a\kb} (\Tra{-\nb}\rb)
							  = \exp{-\I\kb\cdot\Tn} \psi_{a\kb} (\rb).
	\lbl{bloch}
\ee

Now we
proceed to the revised Bloch's theorem, as it was presented in
Ref.~\onlinecite{koskinen_PRL_10}. It holds for a system of electrons
in any external potential invariant under general (isometric) symmetry
operation $\rb'=\Symnb\rb$, that is, when
\be
	V(\Symnb\rb) = V(\rb)
	\lbl{cond1a}
\ee
for any set of commuting symmetry operations $\Sym^{n_i}_i$, $\Symnb \equiv \Sym^{n_1}_1\Sym^{n_2}_2\cdots$.
Using $\Symnb$ instead of $\Tra{\nb}$ in Eq.~\eref{bloch}, 
the revised version of Bloch's theorem, valid for any symmetry,
reads
\be
	\DSymnb \psi_{a\kab}(\rb) = \psi_{a\kab}(\SymInb \rb)
							= \exp{-\I\kaDn} \psi_{a\kab}(\rb).
	\lbl{gbloch}
\ee
Here,
the vector $\kab$ generalizes the conventional $\kb$-vector and the vector of integers $\nb = (n_1,n_2,\dots)$ gives the number of transformations $\Symnb \equiv \Sym^{n_1}_1\Sym^{n_2}_2\cdots$ for each symmetry
$\Sym_i$.
Knowing \wf $\psi_{a\kab}(\rb)$ for arbitrarily chosen simulation cell
is sufficient to calculate the \wf for atoms belonging to any ($\nb^{\text{th}}$) image,
and thereby to solve, again, the electronic structure of the whole system.

Let us next clarify the conditions when the \rblocho~\eref{gbloch} holds.
Because the system should be left unchanged,
the symmetry operations should commute;
therefore, the operators $\DSymY{n_j}{j}$ must commute as well,
\begin{align}
	[\DSymY{n_j}{j}, \DSymY{l_{i}}{i} ] = 0.
	\lbl{comutative}
\end{align}
Condition~\eref{cond1a}
is equivalent to $[V(\rb), \DSymnb] = 0$ and
since $\DSymnb$ commutes also with $\phb^2$,
requirement of potential invariance~\eref{cond1a} becomes
\be
	[\Hh, \DSymnb] = 0.
	\lbl{cond1b}
\ee
The periodic boundary conditions are generalized in the following way. First, write the condition $\psi_{a\kb} (\rb - \Lb_j) = \psi_{a\kb} (\rb)$ in terms of the translation
operator, $\DTra{M_j}\psi_{a\kb}(\rb)=\psi_{a\kb} (\Tra{-M_j}\rb) =
\psi_{a\kb} (\rb)$,
where $M_j = L_j/T_j$ is the number of lattice points along direction $j$. ($\Lb_j$ is the the length of the entire crystal in direction $j$.) Next, replace the symmetry operation $\SymNot T\to\SymNot S$ and $\kb \to \kab$. This gives
$\DSymX{\Mb}\psi_{a\kab}(\rb) = \psi_{a\kab} (\rb)$, or
\be
	\DSymX{\Mb} = \DSymm(\Sym^{M'_1}_1\Sym^{M'_2}_2\dots) = \unity,
	\lbl{cond2}
\ee
where in general $\Mb = (M'_1, M'_2, \dots)$, with $M'_j$ being 0 or $M_j$, where $M_j$ is the number of transformations upon which the system is mapped onto itself. ($\Pi_j M_j=\mathcal{N}$ is the total number of unit cells in the crystal.)

In mathematical terms, Bloch's theorem deals with symmetries of the Hamiltonian alone
and follows from the group representation theory for cyclic groups.
Indeed, considering, for simplicity, the one\B{-}dimensional case, $\nb \to n$,
the set of transformations
\be
	\{
		\unity, \DSymX{1}, \DSymX{2}, \ldots, \DSymX{M - 1}
		\;|\;
		\DSymX{M} = \unity
	\}
	\lbl{group}
\ee
forms a cyclic group
with one-dimensional representation $\exp{\I 2\pi m/M}, m = 0, 1, \ldots, M\!-\!1$.
Due to condition~\eref{cond1b}, this group is also a symmetry group of the Hamiltonian.
Therefore,
the eigenfunctions of the Hamiltonian transform according to representation of its symmetry group,
which is equivalent to Eq.~\eref{gbloch}.

Since any unitary operator can be uniquely represented by an exponent of a hermitian operator \cite{Rose}, 
we have $\DSym = \exp{-\I\kah}$.
Then, $\DSymX{n} = \exp{-\I\kah n}$,
or in multi-dimensional notation
\be
  \DSymX{\nb} = \exp{-\I\kahb\cdot\nb}.
  \lbl{kahb_defined}
\ee
The operator $\kahb$ is commonly called the generator of the group. Due to condition~\eref{cond1b}, $\kahb$ commutes with $\Hh$, and its components $\kah_j$ commute among themselves. Therefore, eigenfunctions of the operator $\kahb$, with $\kahb \Ket{\kab} = \kab \Ket{\kab}$, form a common set of eigenstates with the Hamiltonian operator.
The vector $\kab$ is the eigenvalue of the operator $\kahb$ and
thus a good quantum number that can be used to label the energy eigenstates [as evident already in Eq.\eref{gbloch}]. The physical meaning of vector $\kab$ relies on the symmetry of the system and
is thus specific to given symmetry (see \ssref{examples}).

Now, consider the eigenvalue problem for $\DSymnb$,
\be
	\DSymnb\Kpsi = \exp{-\I\kahDn}\Kpsi = C_{\nb}\Kpsi, 
	\lbl{D-EVP}
\ee
where $C_{\nb}$ is a constant.
We close Eq.~\eref{D-EVP} with $\Bra{\kab}$ to find
\be
	\Bra{\kab}\exp{-\I\kahDn}\Kpsi = \exp{-\I\kaDn} \psi(\kab) = C_{\nb}\psi(\kab),
\ee
yielding $C_{\nb} = \exp{-\I\kaDn}$ with $\psi(\kab) \equiv \BKpsi{\kab}$.
Next we note that it is equivalent to define $\DSymnb$ by
$\psi(\rb) = \psi'(\rb') \equiv \DSymnb\psi(\Symnb\rb)$
and by $\DSymnb \Ket{\rb}  \equiv \Ket{\Symnb \rb}$;
the unitarity property $\DdSymnb = \DISymnb = \DSymnbI$
results in $\Bra{\rb} \DSymnb  = \Bra{\SymInb \rb}$.
Taking this into account,
we
close Eq.~\eref{D-EVP} with $\Brb$ to find
\be
	\Brb\DSymnb\Kpsi = \BKpsi{\SymInb\rb}
	= \exp{-\I\kaDn} \BKpsi{\rb},
\ee
or in coordinate representation
\be
	\DSymnb \psi(\rb) = \psi(\SymInb\rb) = \exp{-\I\kaDn}\psi(\rb).
	\lbl{gbloch-proved}
\ee
Applying the cyclic condition~\eref{cond2},
we arrive at set of allowed values for $\kab$: $\exp{\I\ka_j M_j} = 1 =\exp{\I 2 \pi m_j}  $ or
\be
	\kab =
		2 \pi \left( \frac{m_1}{M_1}, \frac{m_2}{M_2}, \dots \right),
	\lbl{kappas}
\ee
where $m_j = 0,1,\ldots, M_j\!-\!1$.

Finally, labeling the eigenfunction with band index $a$ and the good quantum number $\kab$, Eq.~\eref{gbloch-proved} becomes Eq.~\eref{gbloch}.

We thus proved the \rbloch of Eq.~\eref{gbloch}: for a system of electrons in the external potential resulting from any symmetric arrangement of atoms (or any other external potential for that matter), wave functions separated by $\Symnb$ transformations differ only by a phase factor.
Therefore, by calculating the \wfs in a single unit cell---whatever its form---,
one can simulate the electronic structure of the symmetric system as a whole.

Alternatively, the \rbloch can be formulated by saying that the Hamiltonian eigenfunctions are of the form
\be
	\psi_{a\kab}(\rb) = \exp{\I\kaDn(\rb)} u_{a\kab}(\rb),
	\lbl{gbloch_state}
\ee
where
$u_{a\kab}(\Symmb\rb) = u_{a\kab}(\rb)$ is a periodic function and $\nb(\rb)$ is a generalized dimensionless coordinate, chosen in a way to satisfy
\be
  \nb(\Symmb\rb) = \nb(\rb) + \mb,
  \lbl{n_condition}
\ee
with $\mb = (m_1, m_2, \dots)$ being a vector of integers,
so that \eref{gbloch_state} with condition~\eref{n_condition} would satisfy \eref{gbloch} by construction. The choice of $\nb(\rb)$, however, is not unique. For most transformation $\nb(\rb)$ can be a continuous function, yielding a fraction of the unit transformation for given $\rb$; for example, in one-dimensional case $(\nb \to n)$, at the origin, which also gives the beginning of the simulation cell, $n(\rb) = 0$, at the middle of the simulation cell $n(\rb) = 0.5$, at border between the simulation cell and the first copy of the simulation cell $n(\rb) = 1$. With improper transformation works as well, but requires a discontinuous $\nb(\rb)$.

\begin{figure}
	\center
	\includegraphics[width=\columnwidth]{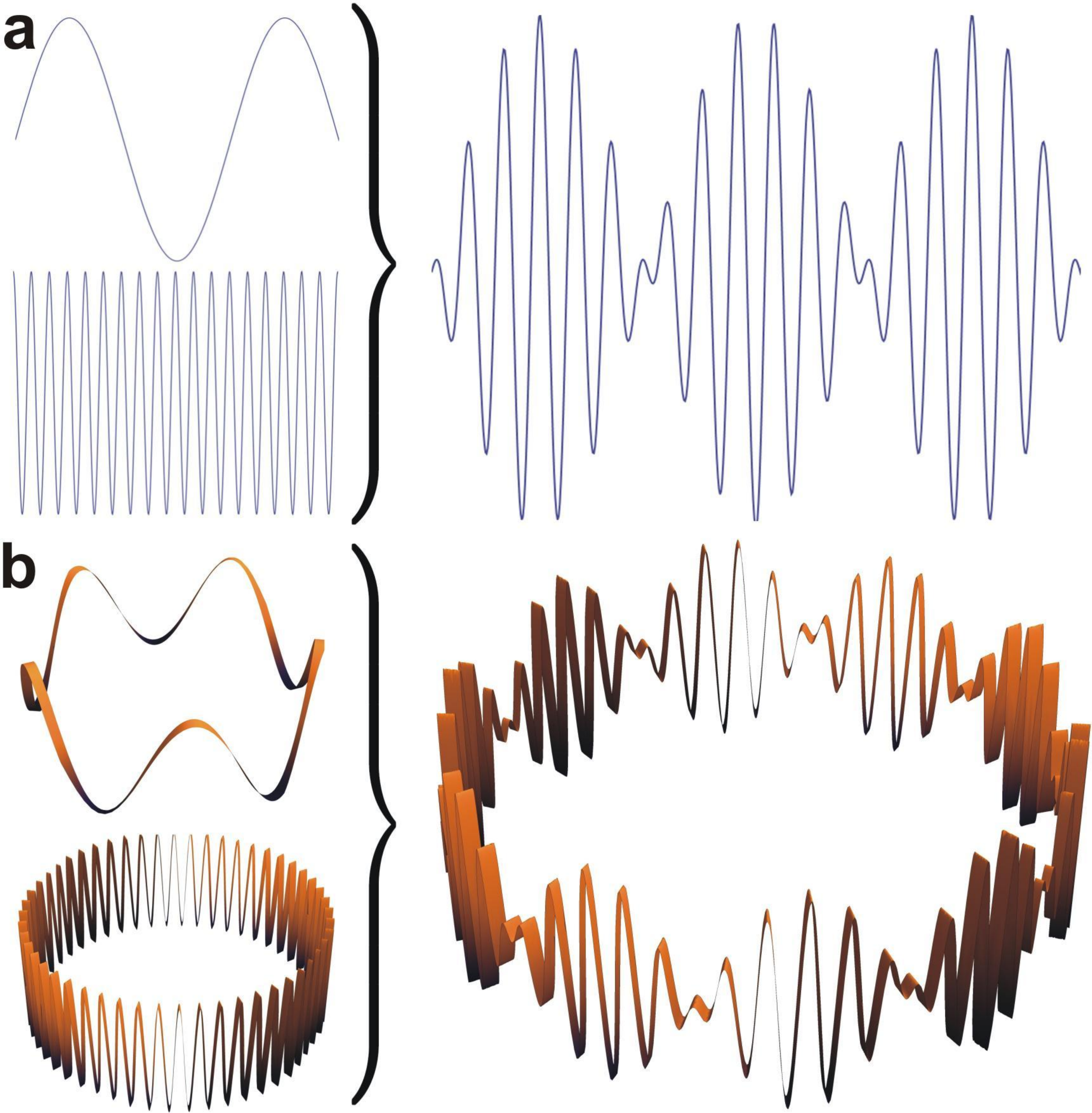}
	\caption{%
	(Color online) \Rblocho:
	(\textbf{a}) The product of the modulating phase factor,
	$\exp{\I\kb\cdot\rb}$,
	and a periodic function $u_{a\kb}(\rb)$
	gives the translational Bloch state.
	(\textbf{b}) The very same principle applies to other symmetries. }
	\label{f-gbloch-state}
\end{figure}

For translations in Cartesian coordinates,
we have
$n_j(\rb) \equiv x_j/T_j$
and
$\ka_j = 2 \pi m_j / M_j$ [Eq.~\eref{kappas}],
which
yields
$\kab \cdot \nb(\rb) = \sum_j 2 \pi m_j / (M_j T_j) \, x_j \equiv \kb \cdot \rb$,
and we
recover
the well-known Bloch state
\be
	\psi_{a\kb}(\rb) = \exp{\I\kb\cdot\rb} u_{a\kb}(\rb),
	\lbl{bstate-transl}
\ee
where $u_{a\kb}(\rb) = u_{a\kb}(\rb + \Tb_{\nb})$
is a lattice-periodic function.
General idea behind formulation~\eref{gbloch_state} is illustrated in \FIG{f-gbloch-state}.
Next, we illustrate this formulation using selected familiar symmetries.

\subsection{Illustrations with familiar symmetries}\label{SS:examples}

A simple transformation to consider is a two-fold rotation ($\pi$-radian rotation).
Since twice repeated it yields the original system
(that is, $\Sym^2 = \unity$, $M = 2$, and $m = 0,1$),
the $\ka$-point sampling is $\ka = 0$ and $\ka = \pi$. Therefore,
\be
	\DSymn \psi_{a\ka}(\rb) = \psi_{a\ka}(\SymIn \rb)
							= (\pm 1)^n \psi_{a\ka}(\rb);
      \lbl{2fold}
\ee
the \wf for $\ka=\pi$ becomes negative upon odd number of transformations.

\def\shortdd{\!:\!}

\setlength{\tabcolsep}{0.15cm}
\begin{table*}
    \caption{
    Generators of one-parameter symmetry transformations.
    Translations are generated by momentum operator $\ph_j$,
    rotations
    by angular momentum operator $\Lh_z$,
    and helical transformations along $z$-axis by $\ph_z$ and $\Lh_z$ together.
    }
    \label{tab:symmetries}
    \centering
        \begin{tabularx}{\textwidth}
{XR@{}L@{\hspace{6ex}}p{2.3cm}L@{\hspace{6ex}}C@{\hspace{5ex}}C}
            \toprule 
            \noalign{\smallskip}
                Symmetry &
                \Symn\shortdd &
                    $~$\rb\to
                    \Symn\rb &&
                \DSymn  &
                 \begin{array}{l}
                        \text{generator}
                    \end{array}
                    &
                \kah_j
            \\
            \midrule 
            \noalign{\smallskip}
                Translational &
                \Tra{n}\shortdd &
                    $~$\rb\to
                    \rb + \Tb_{n}
                    &
                ~~~$\Tb_n = n \Tb_j$
                    &
                \exp{-\I T_{n} \ph_j/\hbar} &
                \ph_j &
                T_j \ph_j/\hbar
            \\\noalign{\smallskip}
                Rotational &
                \Rot{\alb_n}\shortdd &
                    $~$\rb\to
                    \Rot{\alb_n}\rb
                    &
                ~~~$\alb_n = n \alb$
                    &
                \exp{-\I\al_n\Lh_z/\hbar} &
                \Lh_z & 
                \al\Lh_z/\hbar
            \\\noalign{\smallskip}
                Helical &
                \mathcal{X}_{n}\shortdd&
                    $~$\rb\to
                    \Rot{\hib_n} \rb + \Tb_{n}&%
                    $\left\{
                    \begin{array}{l}
                        \hib_n = n \hib
                        \\[0.7ex]
                        \Tb_n = n \Tb_z
                    \end{array}
                    \right.$
                    &%
                    \exp{-\I (T_n \ph_z + \chi_n\Lh_z)/\hbar}%
                    &%
                T_n \hat{p}_z + \chi_n \hat{L}_z &%
                (T_z \ph_z+ \chi\Lh_z)/\hbar
            \\
            \noalign{\smallskip}
            \bottomrule 
        \end{tabularx}
\end{table*}

For $M$-fold rotation around $z$-axis in cylindrical coordinates
$\rb = (\rho, \pfi, z)$,
$\Sym^m (\rho, \pfi, z) = (\rho, \pfi + m\al, z)$, where $\al = 2\pi/M$ is the angle of rotation.
The condition\er{n_condition} is satisfied with $n(\rb) \equiv \pfi/\al$,
and we further take $\ka =  2\pi m/M = \al m$
to write Eq.~\eref{gbloch_state} in terms of azimuthal angle $\pfi$ and quantum number $m$ (which is nothing but the familiar magnetic quantum number):
\be
    \psi_{am}(\rho, \pfi, z) = \exp{\I m\pfi} u_{am}(\rho, \pfi, z),
    \lbl{gbstate_rot}
\ee
where $u_{am}(\rho, \pfi + \al_n, z) = u_{am}(\rho, \pfi, z)$
for any $\al_n \equiv n\al$ $(n = 0,\pm 1,\ldots)$.
For example, with $\al = \pi$, we arrive at \eq{2fold}.
By considering a six-fold rotation symmetry,
we can simulate the 12-atom benzene C$_6$H$_6$
by 2-atom generalized unit cell (one CH unit);
we know the \wf of the whole system if we know the \wf $\psi_{am}(\rho, \pfi, z)$ in the wedge-shaped region $\pfi \in [0, \al]$, $\al = \pi/3$.
On the other hand,
the spatial region for simulation cell does not need to be connected;
in benzene, for example, the CH unit could be constructed from spatial region for C and H even on opposite sides of the molecule.

A helical transformation with pitch length $L_z = M T_z$, is a combination of 
a rotation by an angle $\chi=2\pi/M$ and a translation by $T_z$ where the translation goes along the axis of rotation (here $z$-axis), $\Sym^m (\rho, \pfi, z) = (\rho, \pfi + m\al, z + m T_z)$.
The condition\er{n_condition} can be satisfied with different choices for $n(\rb)$, for example, $n(\rb) = z/T_z$, $n(\rb) = \pfi/\chi$, or $n(\rb) = (z/T_z + \pfi/\chi)/2$, meaning that $z/T_z$, $\pfi/\chi$, or their combination can serve as the dimensionless coordinate.
Further, take $\ka = 2\pi m/M = m\chi$ to write Eq.~\eref{gbloch_state} as
\begin{align}
    \psi_{a{m}}(\rho, \pfi, z)
          &= \exp{\I m (2\pi z/(M T_z) + \pfi){/2}} u_{a{m}}(\rho, \pfi, z),
       \lbl{gbstate_hel}
\end{align}
where $u_{a{m}}$ obeys symmetry via the relation $u_{a{m}}(\rho, \pfi + \al_n, z + T_n) = u_{a{m}}(\rho, \pfi, z)$
for any $T_n = n T_z $ and $\al_n = n \chi$
$(n = 0, \pm 1, \ldots)$.

The above examples of proper symmetry transformations can be described in terms of
the corresponding generators $\kahb$ \tab{tab:symmetries}.
The foundations of the \rblocho, as this section has shown, are very familiar.

\section{The $\kab$-point sampling}

One of the central practical issues in the RPBC approach is the sampling of the $\kab$-values ($\kab$-points). It is essentially similar to the familiar $\kb$-sampling with translational symmetry, although there are some differences. Depending on symmetry, the values of a given component $\kappa_j$ of the $\kab$-vector may or may not be quantized, and sampling of $\kappa_j$ hence falls in two schemes, in either discrete or continuous sampling.

First, in discrete sampling the component $\kappa_j$ accepts only the values given by Eq.~\eref{kappas}, $\left\{2\pi m_j/M_j \,|\, m_j=0,1,\dots,M_j-1\right\}$, where $\ka_j$ correspond to symmetry transformation $\Sym_j$. This usually means a small $M_j$, say $M_j=2$ or $M_j=6$, but even if $M_j$ would be large, say thousand, all thousand values do not necessarily need to be sampled. Non-allowed $\ka_j$'s result in nonphysical wave functions and ultimately problems in numerical evaluations. Discrete sampling suggests that the corresponding symmetry is \emph{genuinely} periodic, and the periodic boundary condition is a real physical condition. The simplest example of this case is the two-fold transformation discussed around Eq.~\eref{2fold}.

Second, in continuous sampling $M_j$ goes to infinity (or can be treated that way), such that $\kappa_j= 2\pi m_j/M_j$-points can be sampled freely between $[0, 2\pi)$, or within the Brillouin zone $[-\pi, \pi)$. Continuous sampling happens for symmetry operations involving translation, for which the periodic boundary is not a real physical condition, but rather a convenient mathematical trick that has turned out useful. Note that in regular three-dimensional system PBC means all dimensions to be periodic in an intertwined and bogus fashion; in two-dimensional system PBC represents topologically a toroid.

Since RPBC works also with translational symmetry, there must be a relation between
the $\kab$-vector and the conventional $\kb$-vector; this was already shown in Ref.~\onlinecite{koskinen_PRL_10}.
It is obtained by equating the exponential factors in Eqs.~\eref{bloch} and \eref{gbloch}, $\exp{-\I\kb\cdot\Tn} = \exp{-\I\kaDn}$. For given $\kb$, by choosing $\nb$ such that only $n_j=1$, we get
\be
	\kab \cdot \nb = \ka_j = \Tb_j \cdot \kb =  \sum_{l=1}^3 T_{j,l} k_{l} \qquad (j = 1,2,3);
	\lbl{k-to-kappa}
\ee
here $l$ runs though Cartesian components of vectors $\Tb_{j}$ and $\kb$. Solving Eq.\eref{k-to-kappa} hence yields $\kab(\kb)$ or vice versa.
It is easy to check that plugging $\kb = \sum_{j=1}^3 m_j/M_j \,\bb_j$ into Eq.~\eref{k-to-kappa} yields the condition~\eref{kappas}, where $\bb_j$ are the reciprocal lattice vectors with $\bb_j \cdot \Tb_{j'} = 2 \pi\delta_{jj'}$. Eq.~\eref{k-to-kappa} suggests also that the $\kab$-points are sampled continuously when $\kb$-point are.

Usually for two symmetries, say $\Sym_i$ and $\Sym_j$, the sampling schemes are independent, meaning that, for example, $\kappa_i$ can have discrete sampling while $\kappa_j$ has continuous sampling. Yet sometimes the symmetry transformations can be coupled, which leads to coupling of $\kappa_i$ and $\kappa_j$ samplings.
This can happen when system is such that
mapping of simulation cell atoms $l_1$ times with $\Sym_{1}$
is identical to
mapping of simulation cell atoms $l_2$ times with $\Sym_{2}$, or $\Sym^{l_1}_1 = \Sym^{l_2}_2$,
which leads to the coupling of $\kab$-points for symmetries $\Sym_1$ and $\Sym_2$.
From the theorem~\eref{gbloch} it follows
that $\exp{-\I\ka_1 l_1} = \exp{-\I\ka_2 l_2}$, and the coupling of the sampling is given by the condition
\be
	l_1\ka_1 = l_2 \ka_2  + 2 \pi m  \qquad (m \text{ integer}).
	\lbl{kacoupling}
\ee
Here values of $\ka_1$ are governed by Eq.~\eref{kacoupling} where $m = 0, 1, \ldots, l_1 - 1$ and $\ka_2$ acts as independent parameter
[alternatively,
$\ka_1$ may act as an independent parameter; then, values of $\ka_2$ would be given by Eq.~\eref{kacoupling} with $m = 0, 1, \ldots, l_2 - 1$].
The sampling of the independent parameter $\ka_2$ [or $\ka_1$] follows either of the two sampling schemes discussed above. In the general case we have $\Sym^{\nb_i}=\Sym^{\nb_j}$, and the couplings of the sampling schemes can be obtained from $e^{-i \kab \cdot \nb_i}=e^{-i \kab \cdot \nb_j + 2\pi m}$.

In the next section, we list cookbook-type recipes how certain symmetries can be used with the RPBC approach. For each symmetry we point out the pertinent $\kab$-point sampling scheme.

\section{Selected Symmetry Setups}

The symmetry of the system determines the coordinate transformations $\Symnb$
and the shape of the simulation cell.
Every particle in the simulation cell, located at $\RI$,
has $\Ncells$ images located at $\RnI \Edef \Symnb \RI$ with $\nb = (n_1, n_2, \ldots)$.
A fairly general symmetry transformation for particles' coordinates
is the affine transformation
\be
				\Symnb \rb \Edef  \Rot{\ombn} \rb + \Tb_{\nb},
	\lbl{coord_transf}
\ee
where
$\Tb_{\nb} \Edef n_1 \Tb_1 + n_2 \Tb_2 + n_3 \Tb_3$
is translation and
$\Rot{\ombn}$ is rotation for an angle $\omn = n_1 \chi_1 + n_2 \chi_2 + n_3 \al$ around an axis given by the vector $\ombn$.
This is not, however, the most general form [\cf
Eqs.~\eref{slab2}--\eref{saddle}].
When rotation is a part of transformation,
there can be at most one translation
that should be collinear with the rotation axis
($\ombn \parallel \Tb_i$).
At the same time, the number of translational transformations along this direction (possibly for different lengths)
is unlimited;
so is the number of coaxial rotational and helical transformations.
Selecting some of the angles $\chi_1$, $\chi_2$, $\al$ and vectors $\Tb_1$, $\Tb_2$, $\Tb_3$ yields the special cases 
we discuss next
(see also \TAB{tab:applications}).
The different symmetries are referred to, from practical simulation viewpoint, as simulation setups. The titles in the following refer to the nomenclature of these setups.

\subsection{Bravais: translational symmetry revisited}
		Transformations of this setup consists of up to three
		translations [$\chi_1, \chi_2, \al \equiv 0$ in Eq.~\eref{coord_transf}]:
		\be
				\Sym^{n_i}_i \rb \Edef  \rb + n_i\Tb_i,\qquad i=1,2,3;
			\lbl{transl}
		\ee
		$\Symnb = \Sym^{n_1}_1 \Sym^{n_2}_2 \Sym^{n_3}_3$.
		This is the typical triclinic setup, mostly used for solids with Bravais lattices. The $\kab$-points, as the $\kb$-points, can be freely sampled. [The direct relation between a given $\kab$-point and a $\kb$-point is given by Eq.\eref{k-to-kappa}].

		Note that, for demonstrative purposes only, one-dimensional translation can also be constructed from two symmetry operations of the form~\eref{transl}, with $\Tb_1 = \Lb_1$ and $\Tb_2 = 2\Lb_1$, $\Symnb = \Sym^{n_1}_1 \Sym^{n_2}_2$. In this case the $\kab$-points are coupled due to the relation $\Sym_1^2 = \Sym_2$. Eq.~\eref{kacoupling} holds, yielding
		$\ka_1 = \ka_2/2 + \pi m$, $M_1 = 2$, and $M_2 \to \infty$.
		This kind of treatment is, of course, silly and artificial, but demonstrates the flexibility of the RPBC approach.

\subsection{Wedge: cylindrical symmetry}
		Main transformation of this setup is rotation
		[$\Tb_2, \Tb_3, \chi_1, \chi_2 \equiv 0$ in Eq.~\eref{coord_transf}; notations of $\Sym_i$ are kept consistent with the discussion following Eq.~\eref{coord_transf}],
		\begin{align}
				\Sym^{n_3}_3 \rb & \Edef  \RotX{n_3\alb} \rb,			
		\end{align}
possibly combined with one translation,
		\begin{align}
				\Sym^{n_1}_1 \rb & \Edef  \rb + n_1 \Tb_{1};
		\end{align}
$\Symnb = \Sym^{n_1}_1 \Sym^{n_3}_3$.
		Translational periodicity in axial direction is either  present
		$(\alb \parallel \Tb_1 \neq 0, M_1 \!\to\! \infty)$,
		or not present
		$(\Tb_1 \equiv 0)$. 
		This setup is for cylindrically-symmetric systems.
		The parameter $\al$ is the angle of the wedge-shaped simulation cell.
		For a large angle $\al$,
		$\ka_3$'s are strictly discrete ($\alpha=2\pi/\text{small integer}$), while for sufficiently small angle $\al$,
		$\ka_3$'s  can be sampled freely
		(the smallness is somewhat subjective, but $\al \lesssim 2\pi/20$ is ``small'' for most practical calculations).

\subsection{Chiral: helical symmetry}
		Transformation of this setup is helical
		[$\al, \chi_2, \Tb_2, \Tb_3 \equiv 0$ in Eq.~\eref{coord_transf}],
		\begin{align}
				\Sym^{n_1}_1 \rb & \Edef  \RotX{n_1\hib_1} \rb + n_1 \Tb_1,
			\lbl{chiral}
		\end{align}
		where rotation and translation are coupled ($\hib_1 \parallel \Tb_1 $).
		This setup is for chiral systems.
		The parameter $\chi_1$ is an angle of shift between two closest chiral units
		at distance $T_z$ apart.
		    The helix makes a full turn in $2\pi/\chi_1$ transformations and its pitch length is $(2\pi /\chi_1)T_z$.
		$\ka_1$ can be sampled freely, since the system is infinitely long ($M_1 \to \infty$).

\begin{table*}[t]
\caption{(Color online) Application of RPBC for distorting materials or improving
  computational efficiency. Some familiar and conventional examples, not
  particularly special for RPBC, are added for completeness. Setup names refer
to the naming scheme used within the \texttt{hotbit} code.\cite{hotbit_wiki,DFTB_Pekka_Ville}}
\begin{tabular}{m{0.54\linewidth}m{0.46\linewidth}}
\toprule 
\noalign{\smallskip}
Illustrations & Examples for structures and distortions \\
%
%
\midrule\noalign{\smallskip} 
\begin{center}
\includegraphics[height=2.7cm]{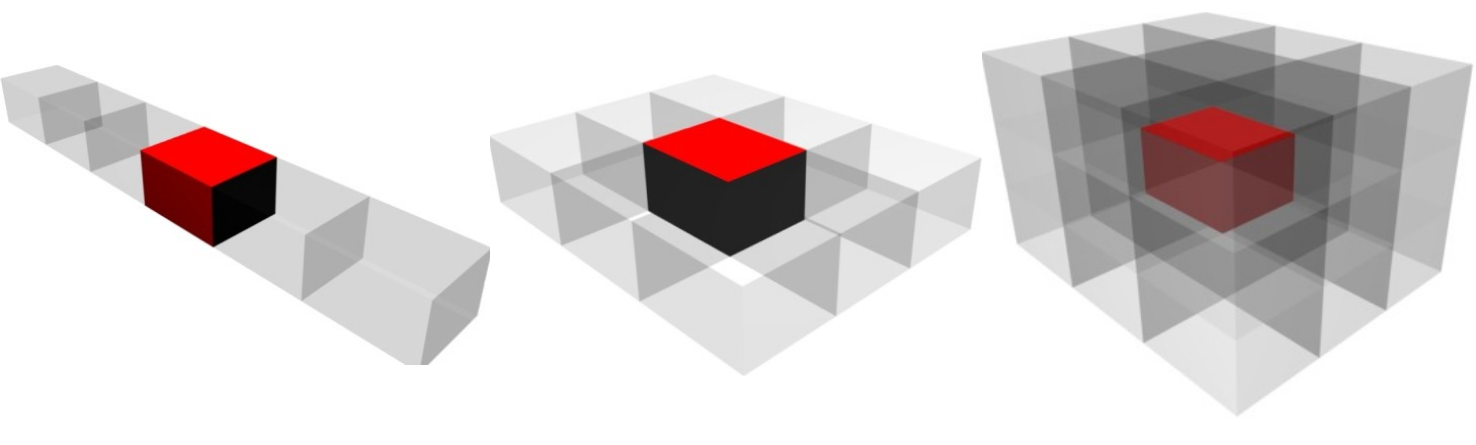}
\end{center} &
Bravais lattices (setup: \texttt{Bravais})
\begin{itemize*}
    \item conventional crystal lattices
    \item translational symmetry in one or more dimensions
\end{itemize*} \\
%
%
\hline\noalign{\smallskip}
\begin{center}
\includegraphics[height=2.7cm]{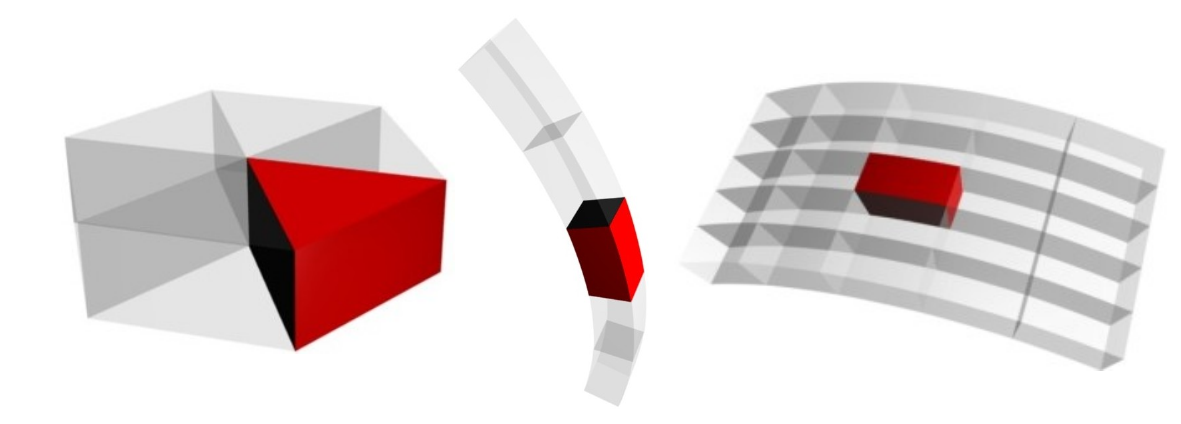}
\end{center} &
Setup: \texttt{Wedge}
\begin{itemize*}
    \item finite, symmetric molecules, clusters, marcomolecules, and viruses with cyclic axes (\eg giant and high genus fullerenes, nanocones, dendrimers)
    \item stretching and bending of one-dimensional structures (nanowires, fibers, rings, tori, rod micelles, polymers, viruses)
    \item bending of planar structures (ribbons, membranes, graphene, thin sheets)
    \item adsorption and catalysis on cylindrical surface
\end{itemize*} \\
%
%
\hline\noalign{\smallskip}
\begin{center}
\includegraphics[height=2.7cm]{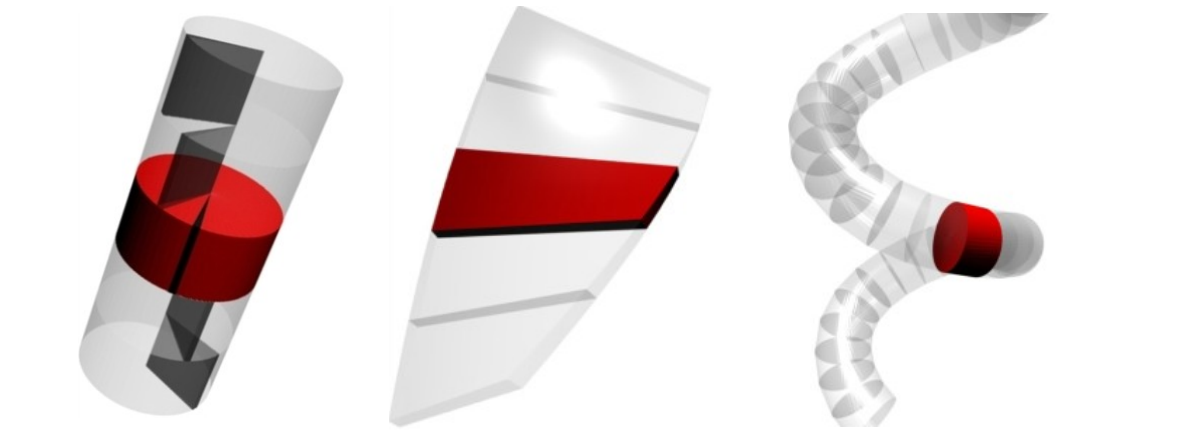}
\end{center} &
Setups: \texttt{Chiral}
\begin{itemize*}
    \item stretching, winding and unwinding of helical structures (chiral and achiral nanotubes, springs, coils, helices, DNA, proteins, polymers, viruses)
    \item twisting of one-dimensional structures (tubes, wires, fibers, ribbons)
\end{itemize*} \\
%
%
\hline\noalign{\smallskip}
\begin{center}
\includegraphics[height=2.7cm]{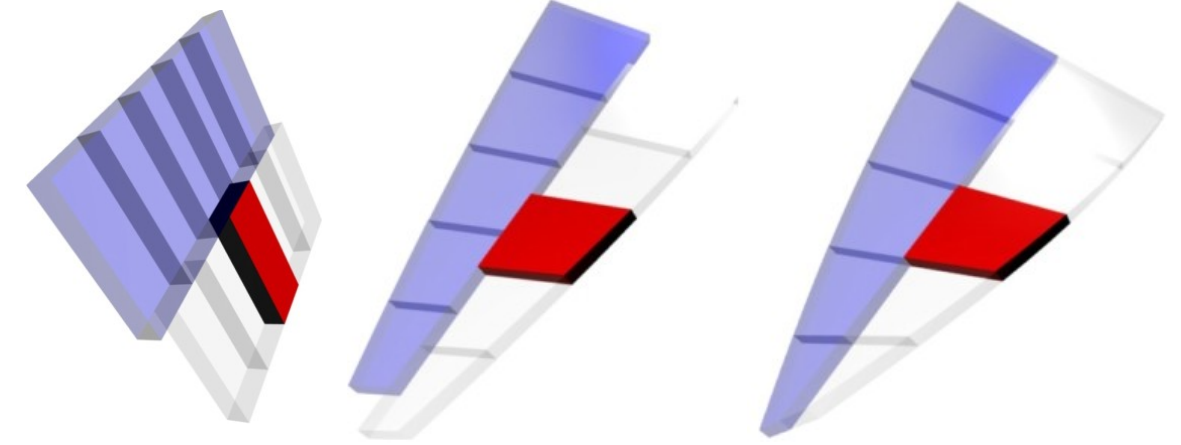}
\end{center} &
Setup: \texttt{DoubleChiral}
\begin{itemize*}
    \item calculating flat or twisted one-dimensional structures with additional (generalized) reflection symmetry
\end{itemize*} \\
%
%
\hline\noalign{\smallskip}
\begin{center}
\includegraphics[height=3.0cm]{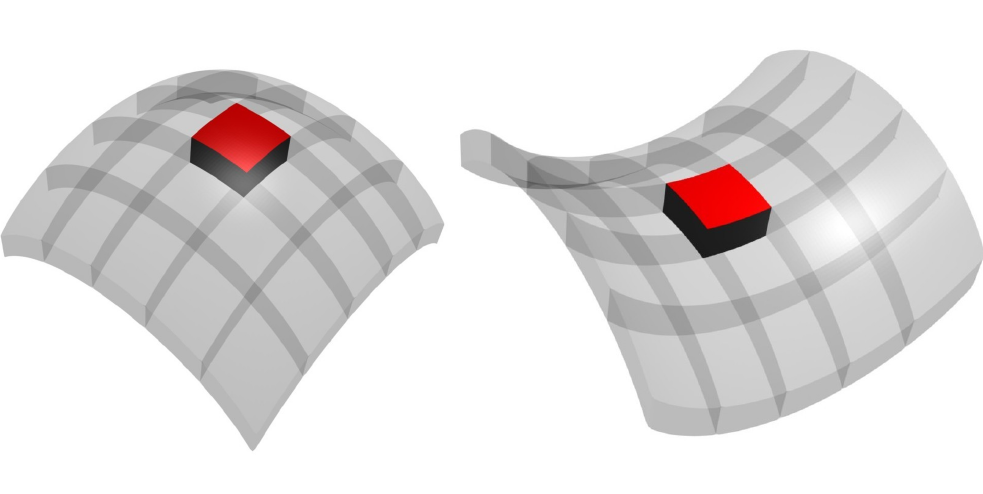}
\end{center} &
Setups: \texttt{Sphere} and \texttt{Saddle}
\begin{itemize*}
    \item approximate treatment only (small curvature)
    \item spherical wrapping of planar structures (graphene, thin films, lipid membranes, vesicles and micelles)
    \item distorting planar structures to a negative Gaussian curvature
    \item adsorption and catalysis on distorted surface
\end{itemize*} \\
%
%
\hline\noalign{\smallskip}
\begin{center}
\includegraphics[height=2.9cm]{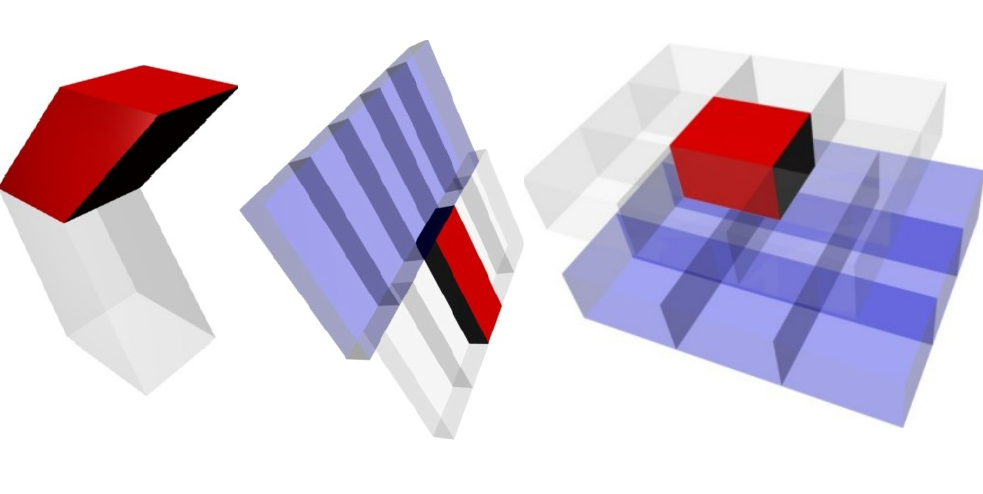}
\end{center} &
Setup: \texttt{Slab}
\begin{itemize*}
    \item one- and two-dimensional wires and slabs with (generalized) reflection symmetry
    \item symmetric finite molecules and clusters
\end{itemize*} \\
\bottomrule 
\end{tabular}
\label{tab:applications}
\end{table*}

\subsection{Double chiral: helical symmetry}
	Helical setup can be extended with second helical transformation [$\al, \Tb_3 \equiv 0$ in Eq.~\eref{coord_transf}]:
	\begin{align}
		\Sym^{n_1}_1 \rb & \Edef  \RotX{n_1\hib_1} \rb + n_1 \Tb_1,
		\\
		\Sym^{n_2}_2 \rb & \Edef  \RotX{n_2\hib_2} \rb + n_2 \Tb_2;
	\end{align}
		both angles $\chi_1,\chi_2$ and both $T_1, T_2$ can be different ($\hib_1 \parallel \hib_2 \parallel \Tb_1 \parallel \Tb_2$).
		This setup is for chiral systems that do not need the whole circumferential part to represent the smallest unit cell.
		Representative examples are carbon nanotubes, where helical transformation have been discussed before\cite{CTWhiteDHRobertsonJWMintmire-PhysRevB.47.5485, lawler_PRB_06, white_JPCB_05, popov_NJP_04}.
		The symmetry implies condition $\Sym^{M_2}_2 = \Sym^{l_1}_1$, and hence
		the restriction $\ka_2 M_2 = \ka_1 l_1 + 2 \pi m $, 
		where either $\ka_1$ or $\ka_2$ is sampled freely.
		Important special case comes from $\chi_2 = \pi$,
		enabling, for example, an efficient simulation of twisted ribbons~\cite{koskinen_PRL_10}.
		Yet another special case arises from  $T_2=0$, where helical transformation is extended by an independent rotation.

\subsection{Slab: surfaces and edges}

	The transformations in this setup are two translations,
	\begin{align}
	      \Sym^{n_1}_1 \rb \Edef \rb + n_1 \Tb_{2},	\\
	      \Sym^{n_2}_2 \rb \Edef \rb + n_2 \Tb_{1},
	\end{align}
and a reflection with partial translations,
	\be
		\Sym^{n_3}_3 \rb \Edef \sigma_{n_3} \rb + \tau_1 n_3 \Tb_{1} + \tau_2 n_3 \Tb_{2},
		\lbl{slab2}
	\ee
where $\tau_1, \tau_2$ are half-integers
and $\sigma_{n_3} \rb \Edef [\rb - 2 (\rb \cdot \zbh) \zbh]^{n_3}$ is a reflection in $xy$-plane.
	This setup is for 2D surfaces and 1D edges ($\Tb_2 = 0$), and allows halving the atom count.
	Values of $(\tau_1, \tau_2)$ are determined by lattice structure of the system in question, being either $(0, 0), (\nicefrac{1}{2}, 0), (0, \nicefrac{1}{2}),$ or $(\nicefrac{1}{2}, \nicefrac{1}{2})$.
	They imply the identity
	$\Sym^2_3 = \Sym^{2\tau_1}_1 \Sym^{2\tau_2}_2$,
	leading to condition
	$\ka_3 = \tau_1 \ka_1 + \tau_2 \ka_2 + \pi m$;
	$\ka_1$ and $\ka_2$ can be sampled freely,
	while sampling of $\ka_3$ is given by $m = 0 , 1$.
	Two values for $\ka_3$ therefore introduce an additional component in two-dimensional band structure plots.

\subsection{Sphere: spherical symmetry (approximate~treatment)}

As discussed in Ref.~\onlinecite{PhysRevB.82.235420}, symmetries can be
treated also in an approximate fashion.
Since for small angles rotations commute to the linear order, $\Rot{\alb_1}\Rot{\alb_2} \approx \Rot{\alb_2}\Rot{\alb_1}$,
two rotations around different axes but same origin, 
\begin{align}
	\Sym^{n_1}_1 \rb \Edef \RotX{n_1\alb_1}\rb,\\
	\Sym^{n_2}_2 \rb \Edef \RotX{n_2\alb_2}\rb,
	\lbl{sphere}
\end{align}
approximately represents a spherical system with simulation cell of a square-conical shape.
The $\kab$-points can be freely sampled.
This setup was recently used to calculate mean and Gaussian curvature moduli of graphene layers~\cite{PhysRevB.82.235420}.

\subsection{Saddle: negative Gaussian curvature (approximate~treatment)}
		Similarly, two rotations around different axes,
		\begin{align}
			\Sym^{n_1}_1 \rb & \Edef \RotX{n_1\alb_1}\rb,	\\
			\Sym^{n_2}_2 \rb & \Edef \RotX{n_2\alb_2}(\rb - \Rb_{0}) + \Rb_{0},
			\lbl{saddle}
		\end{align}
		 such that the origin of first rotation (origin of the coordinate system) and the origin of second rotation, $\Rb_{0}$,
		are located on different sides of the surface,
		approximately represent a saddle system.
		The $\kab$-points can be freely sampled.
		This setup can be used to study membranes with negative Gaussian curvature in approximate fashion~\cite{PhysRevB.82.235420}.

\subsection{On choosing the symmetry setup}

Above all, various symmetry setups introduced by RPBC offer flexible access to different geometries, opening possibilities for distortion studies
(see \TAB{tab:applications}).
Besides, using symmetry setups for symmetric structures---whether classically or quantum-mechanically---bring reduction of simulation costs; atom count reduction is proportional to the number of simulation cell images used in the setup.

Because RPBC approach is centered around the simple coordinate mapping $\rb' \! = \! \Symnb\rb$, computational overhead introduced by RPBC, as compared to translation-periodic implementations, should be in principle small; in a non-orthogonal tight-binding implementation \texttt{hotbit}, the computational overhead turned out negligible.

It is important to choose the setup and the symmetry that fit the system best.
On one hand, too symmetric setup will restrict natural conformations---regarding statistical sampling, for instance, RPBC faces the same artifacts as PBC, should the simulation cell be too small. On the other hand, exploiting too little symmetry will result in unnecessary simulation costs.

Finally, general symmetries may require tighter geometry optimization criteria
in case of small rotational angles. Consider, for example, \FIG{F:forces}b and the total force on the atom in the simulation cell. Although the direct forces due to neighboring images may be large, the small wedge angle projects the combined force to have only a small radial component; geometrical arguments suggest that the radial component is diminished by a factor $\al$ (with $\al$ measured in radians), as compared to the direct forces. In other words, radial forces, although small, may cause certain type of center-of-mass creeping in radial direction, requiring more lengthy and demanding optimization process. With translation symmetry the center-of-mass movements never cost energy.

\section{Towards practice: Warm-up using classical~potential}\label{S:class_pot}

Let us first illustrate the practical aspects of the RPBC formalism with a classical treatment.
Consider the interaction via pair potentials $V_{IJ}(R_{IJ})$,
such that $V_{JI}(R_{IJ}) = V_{IJ}(R_{IJ})$ and $V_{II} \equiv 0$ for all $I$ and $J$,
where $R_{IJ}$ is the length of vector $\Rb_{IJ}=\RJ-\RI$ separating particles $I$ and $J$.
The energy of the entire system, including all $\Ncells$ images of the unit cell, with $N$ particles in each cell, is
\begin{align}
	E' = \half \sum_{IJ} V_{IJ}(R_{IJ}),
	\lbl{pair-pot-extended}
\end{align}
where $I$ and $J$ still run over $K = \Ncells N$ particles.
In order to reduce this expression to a single simulation cell, let $I$ and $J$ run over $N$ particles in one arbitrarily selected cell,
while introducing an index $\nb$ that runs over all its $\Ncells$ images
($\sum_{\nb} 1 = \Ncells$), giving
\begin{align}
	E' & = \half \sum_{IJ}^{N} \sum_{\nb\mb}^{\Ncells} V_{IJ}(|\RmJ - \RnI|).
	\lbl{pair-pot-extended-long}
\end{align}
Then
$\RmJ - \RnI = \Symnb (\RXJ{\mb-\nb} - \RI)
				\Edef \Symnb \RXIJ{\mb-\nb}$,
where $\RnIJ = \RnJ -\RI$ is the vector separating the $\nb^{\text{th}}$ image of particle $J$ and particle $I$. Since transformations $\Symnb$ are isometric,
the energy expression simplifies further to
\begin{align}
	E' = \half \sum_{IJ}^{N} \sum_{\nb\mb}^{\Ncells} V_{IJ}(\RNXIJ{\mb-\nb})
	   =\frac{\Ncells}{2}\sum_{IJ}^{N} \sum_{\nb}^{\Ncells} V_{IJ}(\RNnIJ).\nn
\end{align}

Thus, energy per simulation cell is
\begin{align}
	E = \half \sum_{IJ} \sum_{\nb} V_{IJ}(\RNnIJ).
	\lbl{en_pair_pot}
\end{align}
Here and below, indices $I$ and $J$ run over $N$ particles in the selected simulation cell only; $\nb$ runs over all unit cells where atom $I$ at $\RI$ still interacts with atom $J$ at $\RnJ$.
Comparing energy expressions~\eref{pair-pot-extended-long} and~\eref{en_pair_pot}, one notices the obvious gain in the RPBC approach: Eq.~\eref{en_pair_pot} has one summation less.

To calculate forces as derivatives of energy \wrt particles' positions,
regard $\RnI$ as a function of $\RI$.
In general,
the partial derivatives of $\RnI$ are elements of the rotation matrix $\Rot{\omn}$,
\begin{align}
		\frac{\partial \,[\RnI]_i}{\partial \,[\RI]_j} =  [\Rot{\omn}]_{ij},
		\lbl{partial}
\end{align}
where $i$ and $j$ stand for the Cartesian components.
Starting from $\Fb_{I} = - \nab_I E$, we first arrive at
\begin{align}
	\Fb_{I} =  \half 
											\sum_{J,\, \nb}
												&\{
													 V'_{IJ}(\RNnIJ) \RhnIJ
													 \nl
												&-  V'_{JI}(\RNnJI) \sum_{ij} \left[\RhnJI\right]_{i} [\Rot{\ombn}]_{ij}\ehb_{j}
												\},
	\lbl{beloved_force}
\end{align}
where $V'$ is the derivative of $V$
and $\RhnIJ$ is the unit vector $\RnIJ / \RNnIJ$;
note that $\RhnIJ \neq -\RhnJI$.
The second term in Eq.~\eref{beloved_force}
may seem to be different from the first one, but it is not;
this can be seen by observing the property
\begin{align}
	\RhnJI = - \Rot{\ombn}\RhXIJ{-\nb},
	\lbl{weird_property}
\end{align}
and by using the orthogonality of rotational matrix,
$[\Rot{\ombn}]_{ij} = [\Rot{\om_{-\nb}}]_{ji}$
to simplify the
double sum in the second line of Eq.~\eref{beloved_force} to $-\RhXIJ{-\nb}$.
Thus, the total force acting on ion $I$ in the simulation cell is
\begin{align}
 	\Fb_{I} = \sum_{J} \sum_{\nb}	V'_{IJ}(\RNnIJ) \RhnIJ.
 	\lbl{f_pair_pot}
\end{align}
This expression looks a simple as it should be: the total force exerted on atom $I$ is the sum of the forces from all images of the other atoms $J$.

Note that the expressions~\eref{en_pair_pot} and~\eref{f_pair_pot} are already sufficient to use RPBC in molecular dynamics with classical pair potentials. Generalization to three- or many-body terms is straightforward; a bond-order -type energy expressions, for example, would yield in expressions like
\be
E=\frac{1}{2} \sum_{IJK}^N \sum_{\nb \mb} V_{IJK}(R_{IJ}^{\nb}; \Rb_K^{\mb}).
\ee

\section{Some Peculiarities of Revised PBC}

Differences between orientations of simulation cell images lead to properties worth noting.
Since we are all used to translational setups, these properties appear somewhat peculiar, although their origins are natural.
Although applicable also in quantum mechanical methods, for simplicity we illustrate these properties here using classical pair potentials.

\begin{figure}
	\center

\vspace{1ex}

\includegraphics[width=0.8\columnwidth]{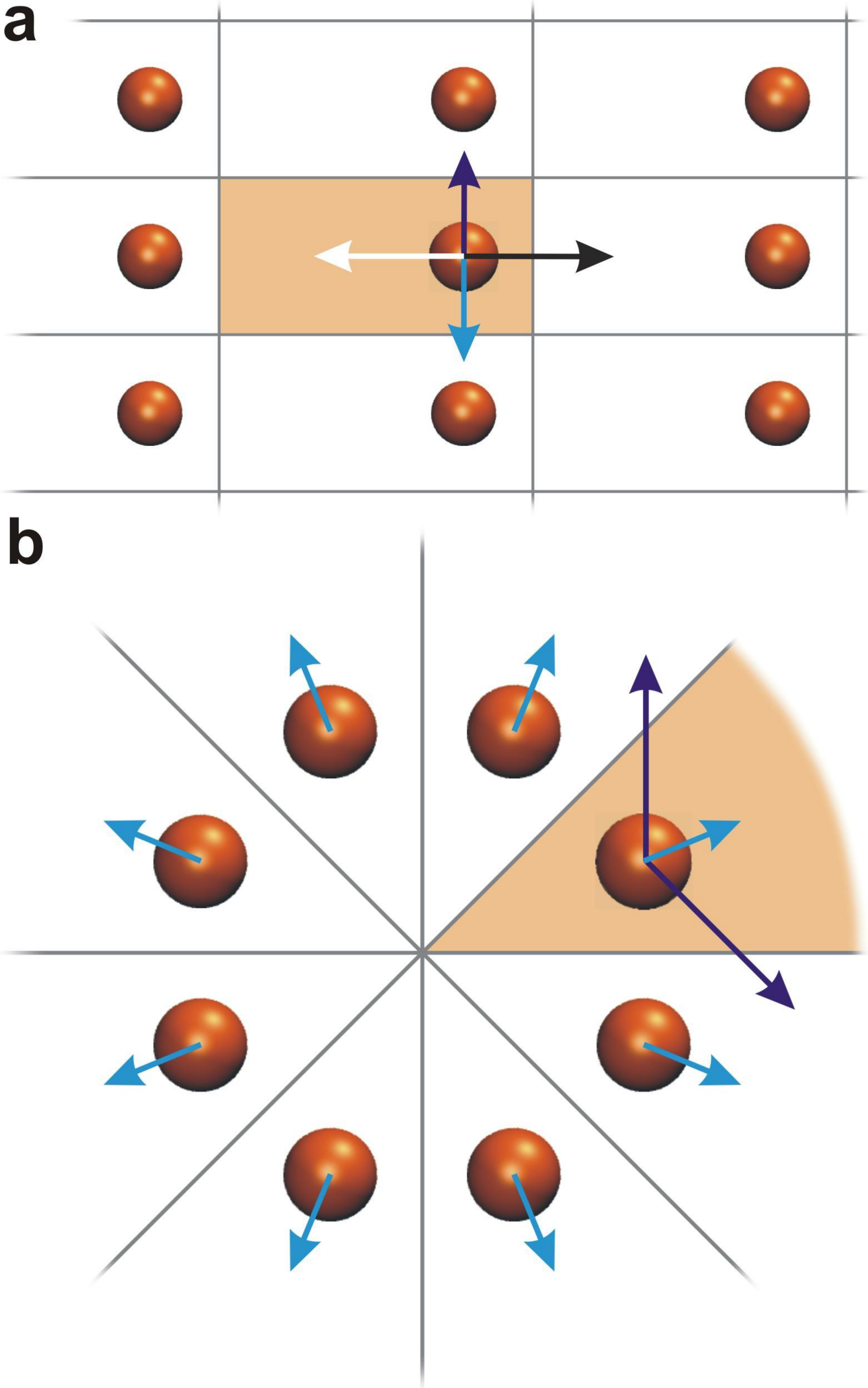}
	\caption{%
	(Color online)
		Peculiarities of symmetry setups illustrated on systems with one particle the in simulation cell (shaded).
		(\textbf{a}) While for translational symmetry  the total force exerted on any atom is zero, (\textbf{b}) in general  the total force (blue or light-gray) can be nonzero:
		a single particle in simulation cell can perform work on itself.
	}
	\label{F:forces}
\end{figure}

First, consider the force exerted on the $\nb^{\text{th}}$ image of atom $I$.
It can be shown to
be symmetric in the sense that
 	$\Fb^{\nb}_{I} = \Symnb \Fb_{I}$
    [Note here that action of
    $\Symnb$ on forces
    goes as for $\rb_{12} = \rb_2 -\rb_1$:
    the operation of $\Symnb$,
    loosely defined for both $\rb_1$ and $\rb_2$ by
    $\Symnb\rb = \Rot{\omn} \rb + \Tb_{\nb}$,
    gives $\Symnb \rb_{12} = \Rot{\omn} \rb_{12}$,
    as $\Tb_{\nb}$'s cancel out].
In the case of translational symmetry, the force exerted on given particle in any cell is the same (lengths and directions of vectors $\Fb^{\nb}_{I}$ and $\Fb_{I}$ are the same).
However, for symmetries that contain rotational transformation, we have
\begin{align}
	\Fb^{\nb}_{I} = \Rot{\omn} \Fb_{I},
	\lbl{force_prop2}
\end{align}
\ie the vector of the force in $\nb^{\text{th}}$ cell is rotated by $\omn$ with respect to the force in simulation cell.
In wedge setup, for example, the forces resulting from cyclic potential
demonstrate cyclic pattern.

Related to this,
consider a system with rotational geometry and one atom in the simulation cell.
Contrary to translations, when the forces exerted on any atom of
a crystal with a monoatomic basis always cancel each other (sum up to zero; see \FIG{F:forces}a),
in the wedge setup it turns out that the resulting force exerted on the atom by its images may differ from zero;
this is evident in \FIG{F:forces}b.
Therefore, a single atom in the simulation cell can perform
work on itself.
In general,
the force exerted on an atom due to its own images is
\begin{align}
 	\fb_{I,I} = \sum_{\nb}	V'_{II}(\RNnII) \RhnII;
	\lbl{partial_self_force}
\end{align}
it may differ from zero
if rotations or reflections are involved.

Further,
consider
a system with wedge geometry and two atoms ($I$ and $J$) in the simulation cell.
Now, the total force exerted on each of these atoms consists of two contributions,
\begin{align}
	\Fb_{I}  & =  \fb_{I,J} + \fb_{I,I},
	\lbl{2at_force1}
\end{align}
and
\begin{align}
	\Fb_{J}  & =  \fb_{J,I} + \fb_{J,J},
	\lbl{2at_force2}
\end{align}
where
\begin{align}
 	\fb_{I,J} = \sum_{\nb}	V'_{IJ}(\RNnIJ) \RhnIJ
 	\lbl{partial_force}
\end{align}
is the force exerted on atom $I$ due to atom $J$ and all the images of $J$,
and
\begin{align}
 	\fb_{J,I} = \sum_{\nb}	V'_{IJ}(\RNnIJ) \RhnJI
 	\lbl{partial_force2}
\end{align}
is the force exerted on atom $J$ due to atom $I$ and all the images of $I$
[note that for a many-atom system, additional terms accounting for all the other atoms would appear in Eqs.~\eref{2at_force1} and~\eref{2at_force2}].
If only translations are involved,
rotation matrix becomes the identity matrix,
Eq.~\eref{weird_property} gives $\RhnJI = - \RhXIJ{-\nb}$, and consequently
	$\fb_{I,J} = - \fb_{J,I}$.
It is tempting to regard this accidental property as Newton's third law.
In general, however,
we can have
$\fb_{I,J} \neq  - \fb_{J,I}$, but notice that
this property is just an artifact of atom indexing.
Namely, by atom $I$ we mean atom $I$ and all its periodic images, \emph{collectively}.
As defined by Eqs.~\eref{partial_force}--\eref{partial_force2}, the forces $\fb_{I,J}$ and $\fb_{J,I}$
are not the forces between two atoms $I$ and $J$ and, therefore,
they do not have to follow Newton's third law in an atom-indexing sense.

\section{Revised PBC with a Practical Quantum Method}\label{S:DFTB}

Here we present the RPBC formulation with a practical quantum-mechanical method; we use a non-orthogonal tight-binding (NOTB) as our method of choice to illustrate the main concepts. To include the main aspects of the present-day standard density-functional (DFT) calculation, we discuss also charge self-consistency and use the concepts from self-consistent charge density-functional tight-binding (DFTB), which has been shown to be a rigorous approximation to full DFT.\cite{elstner_PRB_98} As it will turn out, for an NOTB method RPBC merely introduces a simple transformation for the Hamiltonian and overlap matrix elements. The electrostatic charge fluctuation correction is discussed separately in the next section. The NOTB-related classical repulsion energy has the form of a pair-potential, which was already discussed above. But we start by giving short comments regarding practical basis sets.

\subsection{Comments on basis sets}

Numerically, there are many methods to solve the Schr\"odinger equation\er{shrodinger},
and most of them can be used with RPBC---but some are more practical than others. Real-space grids and especially local basis are particularly suitable for RPBC implementation because they have more freedom to be rotated and adapted to various symmetries. This paper illustrates RPBC with local basis. On the other hand, plane waves are, by their very name, unsuitable for RPBC at least as such.

The plane-wave method
relies on building a basis from plane waves, which are constituted by the factor $\exp{\I\kb\cdot\rb}$ in \eq{bstate-transl}.
The \wf and potential $V(\rb)$ are expanded in a Fourier series, in terms of $e_{\kb}(\rb) \equiv \exp{\I\kb\cdot\rb}$, in order for the Schr\"odinger equation\er{shrodinger} to attain a practical algebraic form.\cite{Marder} Yet Eq.\er{gbloch_state} suggests that a similar type of method could be established also in the general case. Namely, one can build ``symmetry-adapted plane waves'' from the pre-factors $\exp{\I\kaDn(\rb)}$, with function $\nb(\rb)$ chosen such that it properly reflects the symmetry of the system. Once the \wf and potential $V(\rb)$ is expanded in a series with respect to these symmetry-adapted waves $\exp{\I\kaDn(\rb)}$, however, the Schr\"odinger equation\er{shrodinger} not necessarily attain a neat algebraic form. Unraveling these issues are work-in-progress, and it remains a question whether this basis is numerically practical. On the other hand, for analytical investigations Eq.\er{gbloch_state} is useful.

\subsection{Introducing notations: a short primer to NOTB}

The quantum-mechanical method that we use with RPBC is non-orthogonal tight-binding (NOTB) in general, and self-consistent charge density-functional tight-binding (DFTB) in particular.\cite{ElstnerSymposiumIntro, FrauenheimReview, Frauenheim_phys_chem_bio_2000, SeifertReview, DFTB_Pekka_Ville} For establishing notations, we hence review DFTB very briefly; readers familiar with it can proceed directly to the next section.


To represent a single-particle state $\psi_a$,
the DFTB uses a (typically minimal) local basis of $\pfi_\mu$ (LCAO ansatz),
\begin{align}
    \psi_a(\rb)  = \sum_{\mu} c^a_{\mu} \pfi_\mu(\rb),
    \lbl{lcao_expansion}
\end{align}
where
    $\pfi_{\mu}(\rb) \equiv \pfi^0_{\mu}(\rb-\RI)$
is the orbital belonging to the atom $I$ ($\mu \in I$);
$\pfi^0_{\mu}(\rb)$ is localized at the origin,
oriented with respect to fixed Cartesian coordinates.\cite{DFTB_Pekka_Ville}. The total energy is
\begin{align}
    \begin{split}
    E_\text{tot} = & \sum_a f_a \sum_{\mu\nu} c_\mu^{a*}c_\nu^a 
    \widetilde{H}_{\mu\nu} \\
        &  + \half \sum_{IJ} \gamma_{IJ}(R_{IJ})\Delta q_I \Delta q_J
           + \sum_{I<J} V^\text{rep}_{IJ}(R_{IJ}),
    \end{split}
    \lbl{DFTB_energy}
\end{align}
where $f_a$ is the occupation of the state $\psi_a$ and
the Hamiltonian matrix elements
$\widetilde{H}_{\mu\nu} \Edef \Bra{\pfi_\mu} \Hh [n_0] \Ket{\pfi_\nu}$.
(Notation $\widetilde{H}$, instead of the usual $H^0$, is chosen to avoid superscript confusions later.\cite{elstner_PRB_98}) In the second term the excess Mulliken charges $\Delta q_I \Edef q_I - q^0_I$,
as compared to charges $q^0_I$ of valence electrons from neutral system, interact with effective electrostatic interaction $\gamma_{IJ}$.

The efficiency of the DFTB relies on the fact that matrix elements $\widetilde{H}_{\mu\nu}$,
\begin{align}
    \widetilde{H}_{\mu\nu} & = \Itr \pfi^{*}_{\mu}(\rb) \Hh \pfi_{\nu}(\rb)                         \nl
                 & = \Itr \pfi^{0*}_{\mu}(\rb - \Rb_I) \Hh \pfi^0_{\nu}(\rb- \Rb_J)
        \lbl{H0munu}
\end{align}
with $\mu \in I, \nu \in J$, as well as $S_{\mu\nu}$,
are tabulated with respect to distances between orbital centers $\RNIJ = |\Rb_J - \Rb_I|$. The matrix elements with all the possible orientations of the orbitals $\pfi^{0*}_{\mu}$ and $\pfi^0_{\nu}$
are accounted for by a superposition of the matrix elements with Slater-Koster transformation tables~\cite{slater_PR_54} and just a few basic matrix elements.

\subsection{Revised Bloch waves}

The Bloch basis functions for translational symmetry, or Bloch waves, are given with a local basis $\varphi_\mu(\rb)$ by the familiar expression \cite{DFTB_Pekka_Ville}
\begin{align}
	\pfi_{\mu\kb}(\rb) & \Edef
			\oOrn \sum_{\Tb_{\nb}}\exp{\I\kb \cdot \Tb_{\nb}}
					\Tra{\nb}\pfi_{\mu}(\rb),
	\lbl{bloch_basis}
\end{align}
where $\Tra{\nb}\pfi_{\mu}(\rb)  = \pfi_{\mu}(\rb-\Tb_{\nb})$;
Bloch waves extend through the entire system.
Analogously,
for general symmetries,
atomic orbitals of all $\Ncells$ images,
\be
	\pfi^{\nb}_{\mu}(\rb) \Edef \DSymnb \pfi_{\mu}(\rb),
	\lbl{nth_orbital}
\ee
are assembled to create new basis functions according to
\begin{align}
	\pfi_{\mu\kab}(\rb) & \Edef
						\oOrn \sum_{\nb}\exp{\I\kaDn} \DSymnb \pfi_{\mu}(\rb),
	\lbl{gbloch_basis}
\end{align}
with $\pfi_{\mu\kab}(\rb) \Edef \BK{\rb}{\pfi^{\kab}_{\mu}}$.

Single-particle states
\be
	\psi_{a\kab}(\rb) \Edef \sum_{\mu} c^a_{\mu}(\kab) \pfi_{\mu\kab}(\rb)
	\lbl{ks_exp}
\ee
then fulfill the \rblocho, Eq.~\eref{gbloch}, by construction.

\begin{figure}
	\center
	\includegraphics[width=0.8\columnwidth]{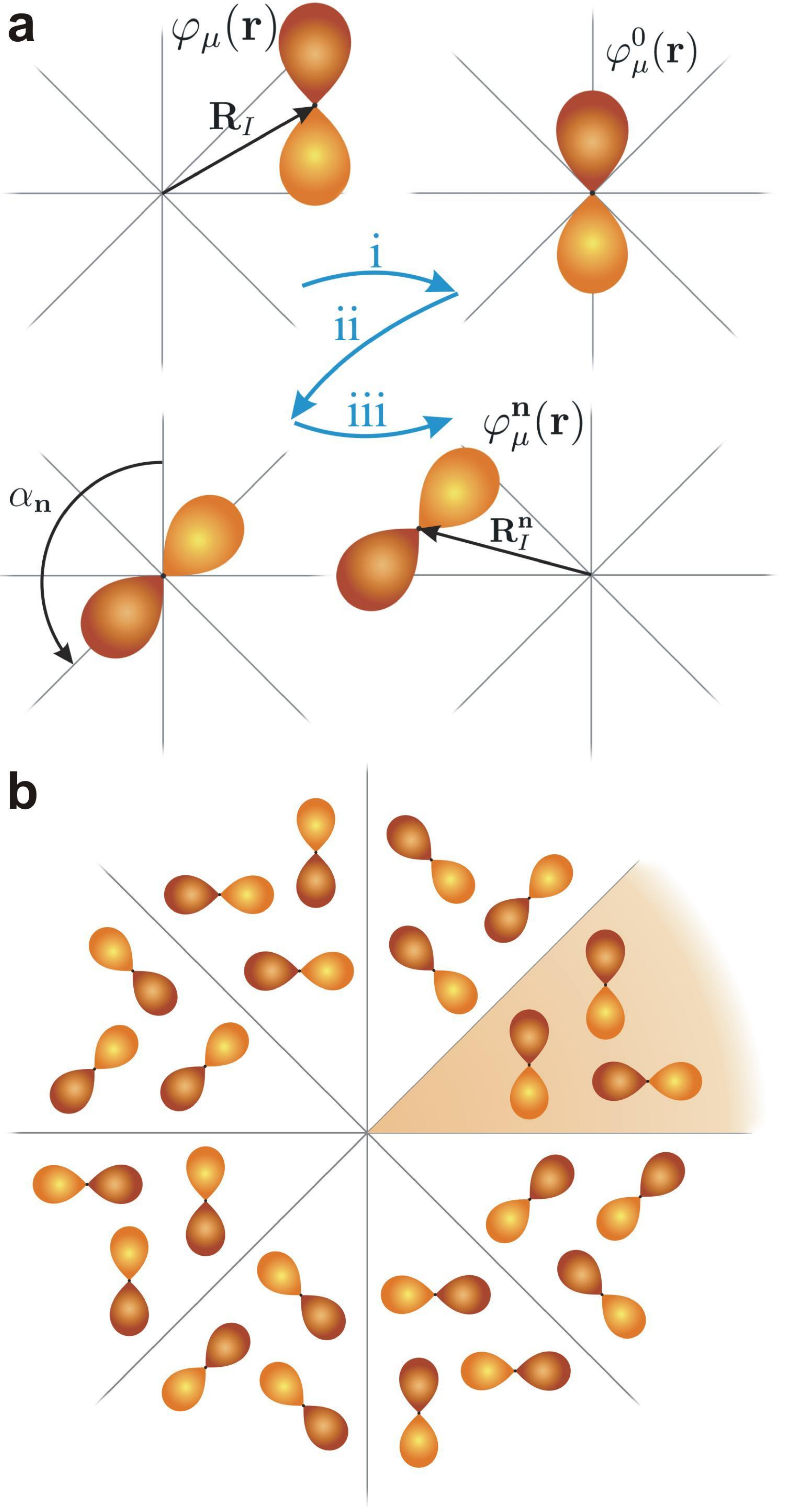}
	\caption{%
	(Color online)
	(\textbf{a}) Illustration of Eq.~\eref{DSSymnb}: three-step transformation of orbital $\pfi_{\mu}(\rb)$ to orbital $\pfi^{\nb}_\mu(\rb)$.
	(\textbf{b}) Illustration of orbitals $\pfi^{\nb}_\mu(\rb)$ for the entire system in wedge setup. Orbitals are symmetrically aligned (simulation cell is shaded).
	}
	\label{F:orbitals}
\end{figure}

The image orbital $\pfi^{\nb}_{\mu}(\rb)$
is the simulation cell orbital $\pfi_{\mu}(\rb)$ transformed by $\DSymnb$, and is not an eigenfunction of $\DSymnb$.
When acting on the orbital $\pfi_{\mu}(\rb)$, located on atom $I$ at $\RI$ and oriented along Cartesian axes, transformation $\DSymnb$ can be thought to be the superposition of three consecutive transformations of
i) translation of the orbital from atom $I$ to origin [becoming hence the orbital $\pfi^0_{\mu}(\rb)$],
ii) rotation of the orbital though $\aln$-radians, to adapt to the ``orientation'' of the image $\nb$,
and
iii) translation of the orbital to $\RnI$, the position of atom $I$ in the image $\nb$ 
(\FIG{F:orbitals}a), \ieo,
\be
	\DSSymnb = \DTTra{\RnA} \DRRot{\aln} \DTdTra{\RA}.
	\lbl{DSSymnb}
\ee
This gives
\begin{align}
	\pfi^{\nb}_\mu(\rb) & = \sum_{\mu'}
		\DRRotM{\aln}{\mu'\mu} \pfi_{\mu'}(\rb-(\Rnmup -\Rmup)) ,
	\lbl{phi_n}
\end{align}
where
$\pfi_{\mu'}(\rb-(\Rnmup -\Rmup)) = \pfi^0_{\mu'}(\rb - \Rnmup)$,
with $\Rb_\mu$ meaning $\Rb_I$ such that $\mu \in I$, and
\begin{align}
	\DRRotM{\aln}{\mu\nu} \Edef \Bra{\pfi^0_\mu} \DRRot{\aln} \Ket{\pfi^0_\nu}
	\lbl{rep_def}
\end{align}
being the representation of rotation $\DRRot{\aln}$ in basis of orbitals
$\pfi^0_{\mu}(\rb)$ located at the origin with the conventional orientation along Cartesian axes;
the explicit form for $\DRRotM{\aln}{\mu\nu}$ is given in Appendix~\ref{S:trmatrix}.
The orientation of the orbitals $\pfi^{\nb}_\mu(\rb)$ in different images $\nb$
hence follow the symmetry,
as illustrated in \FIG{F:orbitals}b.

\subsection{Transformation of matrix elements}

In the revised Bloch basis, Eq.~\eref{gbloch_basis}, the Hamiltonian and overlap matrix elements
$H_{\mu\nu}(\kab,\kab') \Edef \Bra{\pfi^{\kab}_\mu} \Hh \Ket{\pfi^{\kab'}_\nu}$
and
$S_{\mu\nu}(\kab,\kab') \Edef \BK{\pfi^{\kab}_\mu}{\pfi^{\kab'}_\nu}$
transform as well;
we need to account for orbital rotations.

While here we consider Hamiltonian matrix elements only, similar expressions hold for overlap matrix elements as well. Using the explicit form~\eref{gbloch_basis} for new basis functions, we obtain
\begin{align}
	H_{\mu\nu}(\kab,\kab')
			& \Edef \Bra{\pfi^{\kab}_\mu} \Hh \Ket{\pfi^{\kab'}_\nu}		\nl
			& = \oOn \sum_{\mb,\nb} \exp{-\I\kaDm+i\kab'\cdot\nb}
									H_{\mu\nu}(\mb, \nb),			\nn
\end{align}
where
\be
	H_{\mu\nu}(\mb, \nb) \Edef
		\BfiiX{\mb}{\mu} \Hh \KfiiX{\nb}{\nu}
	\lbl{reduced_matrix_el_full}
\ee
($H$ and $\widetilde{H}$ transform the same way, so we use $H$ instead of $\widetilde{H}$).
Since $\Hh$ and $\DSymnb$ commute, $\BfiiX{\mb}{\mu} \Hh \KfiiX{\nb}{\nu} = \BfiiX{}{\mu} \Hh \KfiiX{\nb-\mb}{\nu}$,
or with a shorter notation
\begin{align}
	 H_{\mu\nu}(\mb,\nb) = H_{\mu\nu}(0, \nb-\mb) \Edef H_{\mu\nu}(\nb-\mb).
	\lbl{reduced_matrix_el}
\end{align}
Hence the matrix elements become diagonal in $\kab$, as expected,
\begin{align}
    H_{\mu\nu}(\kab) \Edef \sum_{\nb}\exp{\I\kaDn} H_{\mu\nu}(\nb)\lbl{ka_mat_el_short},
\end{align}
and the $\kab$-dependence of matrix elements ends up resembling translation-periodic formalism\cite{DFTB_Pekka_Ville} precisely---only $\kb$ has changed to $\kab$.

However, the matrix element, Eq.~\eref{reduced_matrix_el}, is now given by
\begin{align}
	H_{\mu\nu}(\nb)
		& = \Itr\fii{*}{\mu}(\rb) \Hh  \left[ \DSymnb \pfi_{\nu}(\rb) \right],
	\lbl{reduced_matrix_element1}
\end{align}
and becomes hence more complex due to rotations induced by $\DSymnb$. By substituting Eq.~\eref{phi_n} inside the square brackets, a short manipulation yields
\be
	H_{\mu\nu}(\nb)
	  = \sum_{\nu'} H'_{\mu\nu'}(\nb) \DRRotM{\alb_{\nb}}{\nu'\nu},
	\lbl{Hn}
\ee
where
\begin{align}
	H'_{\mu\nu}(\nb)
		&\Edef \Itr \pfi^{*}_{\mu}(\rb) \Hh
						\pfi_{\nu}(\rb -(\Rnnu-\Rb_{\nu}))	\nl
		&= \Itr \pfi^{0*}_{\mu}(\rb - \Rmu)
				 			\Hh \pfi^0_{\nu}(\rb - \Rnnu)
	\lbl{H_shifted}
\end{align}
is the matrix element involving translations only, with orbitals $\pfi_{\mu}$, centered at $\Rmu$, orbitals $\pfi^0_{\mu}$, centered at the origin,
all orbitals having a \emph{fixed} orientation. (Orbitals oriented \wrt the fixed Cartesian coordinates of the simulation cell.)
The matrix element $H'_{\mu\nu}$ as given by Eq.~\eref{H_shifted} is
practically the same as the one in
Eq.~\eref{H0munu},
and it is straightforward to calculate in any tight-binding implementation.
Eq.~\eref{Hn} is a simple matrix multiplication, and in compact matrix notation
\be
	\HM(\nb) = \HpM(\nb) \cdot \DRotMM{\aln}.
	\lbl{HnM}
\ee
Thus, compared to the translation-periodic formalism  \cite{DFTB_Pekka_Ville}, matrix elements need simply to be transformed by $\DRotMM{\aln}$.
Eq.~\eref{HnM} is valid within the two-center approximation, otherwise one needs to return to Eq.~\eref{reduced_matrix_element1}.

\subsection{Total energy and forces in NOTB}

The total energy expression for NOTB with revised PBC is therefore
\begin{align}
		E_\text{tot}  = & \sum_{\kab} w_{\kab} \Tr{\rhoM(\kab) \cdot \widetilde{\HM}(\kab)}	\nl
						&+ E_\text{2nd}
						+ \sum_{I<J} V^\text{rep}_{IJ}(R_{IJ}),
		\lbl{ka_energy}
\end{align}
where
$w_{\kab}$ are weights for equivalent $\kab$-points
$(\sum_{\kab} w_{\kab}=1)$, and
\begin{align}
		\rho_{\nu\mu}(\kab) =
							\sum_{a}  f_a(\kab) %
						\left[ 	c^{a*}_{\mu}(\kab) c^a_{\nu}(\kab)	\right]
\end{align}
is the $\kab$-dependent density matrix.
Mulliken charge analysis is done as before (although with transformed $S$-matrix),
and
the electrostatic charge-fluctuation correction remains the same,
\be
	E_\text{2nd} = \half \sum_{IJ} G_{IJ} \Delta q_I \Delta q_J,
	\lbl{E2nd}
\ee
except that the expression
\be
	G_{IJ} = \sum_{\nb} \gamma_{IJ}(\RNnIJ)
	\lbl{elstat}
\ee
is different, now using the generalized symmetry.

Minimizing
$E_\text{tot} - \sum_a f_a(\kab) \eps_a(\kab) \left[\BK{\psi_a}{\psi_a} -1\right]$
\wrt $c^{a*}_{\mu}(\kab)$
yields the generalized eigenvalue problem
\begin{align}
	\sum_{\nu} c^a_{\nu}(\kab) [H_{\mu\nu}(\kab) - \eps_a(\kab) S_{\mu\nu}(\kab)] = 0,
			\quad\forall \kab, \mu, a,
	\lbl{ka_secular}
\end{align}
where
\begin{align}
	H_{\mu\nu}(\kab)  \Edef  \widetilde{H}_{\mu\nu}(\kab) + S_{\mu\nu}(\kab) \sum_K
			\half (G_{IK} + G_{JK}) \Delta q_K \nn
\end{align}
for $\mu \in I$ and $\nu \in J$.
The secular Eq.~\eref{ka_secular} looks very familiar---RPBC does not alter the appearance of the NOTB formalism at all.

Differentiation of the band structure energy \wrt the atoms' positions yields
band-structure forces straightforwardly as
\begin{align}
	 \Fb_I & = - \sum_{\kab} w_{\kab} \Trp{I}{
	 		\db\Hb(\kab)\rhob(\kab) - \db\Sb(\kab)\rhob^\text{en}(\kab)
	 	}.
	 \lbl{ka_force_implement}
\end{align}
Here $\TrX_I$ stands for partial trace over orbitals of atom $I$ only,
\be
	\rho^\text{en}_{\nu\mu}(\kab) \Edef \sum_{a} f_a(\kab) \eps_a(\kab)
								\left[
									c^{a*}_{\mu}(\kab) c^a_{\nu}(\kab)
								\right]
	\lbl{e_den_matrix}
\ee
is the energy-weighted density matrix,
and
\begin{align}
	 \db H_{\mu\nu}(\kab) \Edef  - \sum_{\nb}\exp{\I\kaDn} \nab_J H_{\mu\nu}(\nb), \quad \nu \in J
\end{align}
with the same for $\db\Sb(\kab)$.
Gradients $\nab_J H_{\mu\nu}(\nb)$, $\nab_J S_{\mu\nu}(\nb)$ can be straightforwardly calculated from
the gradients of $H'_{\mu\nu}(\nb)$ and $S'_{\mu\nu}(\nb)$, Eqs.~\eref{Hn}--\eref{H_shifted}.

We still wish to emphasize the obvious result: while internal structure of the Hamiltonian and overlap matrix elements $H_{\mu\nu}(\kab)$ and $S_{\mu\nu}(\kab)$ for general symmetries are different from the translational case, the final equations are trivially the same---$\kb$ only changes to $\kab$.

\section{Electrostatics}

\def\v{\mathbf}
\def\n{\hat}
\def\t{\mathbf} 
\def\tr{\text{tr}}
\def\sgn{\mathop{\rm sgn}}

A crucial component of the interatomic interaction in ionic or partially
polarized solids is electrostatics.
Na\"ive direct summation of the long-ranged electrostatic pair interaction,
while possible using the approach outlined in Sec.~\ref{S:class_pot}, is
slow and only conditionally convergent and hence numerically prohibited.

Methods with better convergence properties have been established for periodic
systems with translational symmetry in three dimensions.
Most of these are based on Ewald summation~\cite{Toukmaji:1996p73} which do
not straightforwardly generalize to partially periodic systems.
Hence, we use a direct summation method whose convergence is accelerated by
the use of a hierarchically telescoped multipole
expansion~\cite{Lambert:1996p274}, in the spirit of the Fast Multipole
Method (FMM)~\cite{Greengard:1994p909,Greengard:1987p325,Rokhlin:1985p187}.
In a multipole expansion scheme, the charge distribution $\rho$ and
electrostatic potential $\Phi$ are expanded around suitably chosen
origins.
These expansions are called the multipole expansion $M$ and the local
expansion $L$, respectively.
We will carry out the expansion in terms of spherical harmonics, but also
Taylor expansions have been reported \cite{Ding:1992p4309}.

\begin{figure}
  \center
  \includegraphics[width=8.5cm]{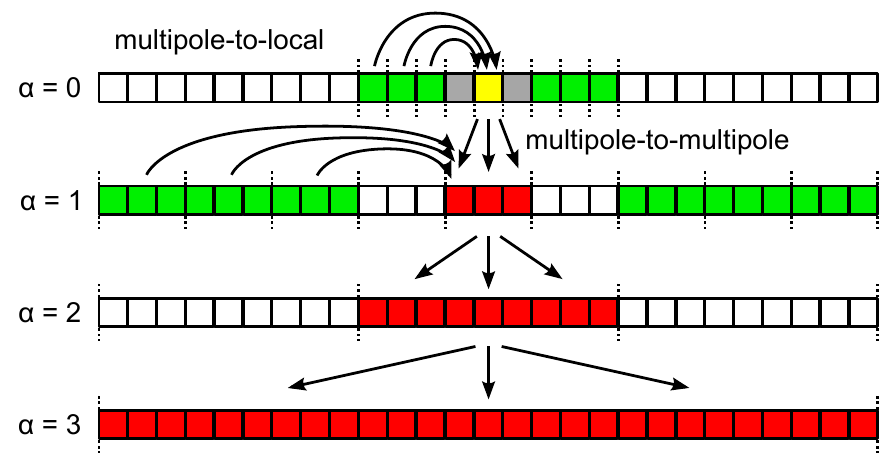}
  \caption{%
    (Color online) Sketch of the operations needed for the telescoped
    fast-multipole expansion demonstrated for an example one-dimensional
    Bravais lattice and $N_1=3$. The value of $\alpha$ denotes the level of
    the expansion. The
    method computes the electrostatic potential and field within the
    simulation cell (yellow). {\bf multipole-to-multipole operation:} At
    $\alpha=0$, the multipole moment $\t{M}_0$ is the moment of the simulation cell
    (yellow). At each successive level $\t{M}_\alpha$ is the combined multipole
    moment of $N$ (here $N=3$, red) units of level $\alpha-1$, where the basic
    repeat unit at level $\alpha$ is denoted by the broken lines. {\bf
      multipole-to-local operation:} At each level $\alpha$, the local expansion at
    the origin is determined from the multipole expansion at level $\alpha$ (from
    the green cells). {\bf local-to-local operation:} Within the simulation
    cell (yellow), the local expansion is shifted to each particle to
    determine the electrostatic potential and field at the particle's
    position. This is then combined with the near-field contribution (from the
    gray repeat units).
    }
  \label{F:fmp}
\end{figure}

The determination of the electrostatic potential and field requires three
operations (see also Figure~\ref{F:fmp}).
First operation is to shift the origin of the multipole expansion; it is
commonly called the multipole-to-multipole transformation.
This operation is required to telescope the multipoles, \ie to join a number
of adjacent multipole expansions into a single one.
Second operation is to determine the local expansion from the
multipole expansion, the multipole-to-local transformation.
This operation solves the electrostatic problem and is typically the most
computationally demanding step.
The third and final operation is to shift the origin of the local expansion to
arbitrary points in space, the local-to-local transformation.

Since the method operates in real space, RPBCs are straightforwardly
implemented in such a scheme, because it requires only the mapping $\rb' =
\Symnb \rb$.
Basically, in addition to translation (\ie the multipole-to-multipole
transformation) the multipole moments of the expansion need to be rotated.
For spherical harmonics, such a rotation is described by the Wigner
$\t{d}$-matrix~\cite{weissbluth} and hence the rotation operation is similar
to the rotation of the orbitals described in section \ref{S:DFTB}.
The general scheme for electrostatics in RPBCs, that includes reflection and
inflection operations, is described in more detail in the following.
Figure~\ref{F:fmp} shows a graphical illustration of the individual operations.

We would like to note that such multipole expansion scheme is also
straightforwardly transferable to full DFT with Gaussian or other
atom-centered basis functions.
An example of such a method for nonperiodic systems is the Gaussian very fast
multipole method (GvFMM)~\cite{Burant:1996p43,Strain:1996p51} and the
continuous fast multipole method (CFMM)~\cite{White:1996p268}.
Since in the far-field the only relevant moment of the Gaussian charge
densities is the monopole, the far-field solution for Gaussians is identical
to that obtained in the FMM for point charges.
The near field solution contains two-electron repulsion integrals which need
to be evaluated directly~\cite{White:1996p2620}.
Indeed, the $\gamma_{IJ}$ Coulomb integral of the self-consistent charge NOTB
formalism is the two-center expression of such an overlap contribution for the
interaction of Gaussian charge distributions with s-symmetry.

\subsection{Multipole expansion}

The multipole moments are given by
\begin{equation}
  M_l^m = \sum\limits_I \Delta q_I R_l^m(\Delta \v{R}_I)
  \label{multipoles}
\end{equation}
where $I$ denotes atoms and $\Delta \v{R}_I$ is the position vector of atom $I$ computed with respect to some expansion origin $\v{R}_0$, \ie $\Delta \v{R}_I=\v{R}_I-\v{R}_0$. Note that $|m| \leq l$. The $R_l^m(\v{r})$ in Eq.~(\ref{multipoles}) are the regular solid harmonics,
\begin{equation}
  R_l^m(\v{r}) = \frac{1}{(l+m)!} r^l e^{-i m \varphi} P_l^m(\cos\theta),
  \label{solid_harmonics_R}
\end{equation}
where $P_l^m(x)$ are the associated Legendre polynomials and $r$, $\theta$ and $\varphi$ are the length, inclination and azimuth angle of $\v{r}$, respectively. The corresponding conjugates are the irregular solid harmonics $I_l^m(\v{r})$ given by
\begin{equation}
  I_l^m(\v{r}) = (l-m)! r^{-(l+1)} e^{i m \varphi} P_l^m(\cos\theta).
  \label{solid_harmonics_I}
\end{equation}
The expansion of the electrostatic interaction into multipoles is based on the
identity~\cite{Jackson:Book1999}
\begin{equation}
  \frac{1}{|\v{r}-\v{r}'|}
  =
  \sum\limits_{\lambda=0}^\infty
  \sum\limits_{\mu=-\lambda}^\lambda
  R_l^m(\v{r}) I_l^m(\v{r}')
  \label{multipl}
\end{equation}
for the free-space Green's functions. The overall scheme is hence a real-space
method. A multiplicative decomposition of the kernel such as Eq.~(\ref{multipl}) can
also be achieved using wavelets~\cite{Harrison:2004p11587,Alpert:1993p246}
that are indeed related to the fast multipole method~\cite{Beylkin:1991p141}.
It is hence conceivable that wavelet-based algorithms could be used here
instead of the currently proposed scheme.

Note that the normalization used in Eqs.~(\ref{solid_harmonics_R}) and (\ref{solid_harmonics_I}) is non-standard. The usual solid harmonics are given by $\sqrt{(l-m)!(l+m)!} R_l^m(\v{r})$. This particular choice of normalization is motivated by numerical considerations. At large expansion orders $l$ the value of $r^l$ becomes large and the factor $(l+m)!$ provides some compensation. While this is not a rigorous approach towards numerical stability, in all practical situations the expansion is cut-off at a value $l_\text{max}$. The approach outlined here is stable up to at least $l_\text{max}=10$. Well-converged solutions are typically obtained for $l_\text{max}=8$.

\subsection{Multipole-to-multipole, multipole-to-local, and local-to-local transformations}

The general transformations underlying the fast-multipole expansion method have been described in detail elsewhere~\cite{White:1994p6593,Greengard:1987p325} and can indeed be found in textbooks on electricity and magnetism~\cite{Jackson:Book1999}. We will therefore only summarize the basic equations, and indicate where the symmetry transformation operators $\Symnb$ need to be applied.

\emph{Multipole-to-multipole.} Let $\v{R}_0$ denote the origin of the
multipole expansion $\t{M}$. The expansion $\t{\tilde M}$ is located at $\v{0}$. Then, the expansion coefficients $M_l^m$ are given by the convolution
\begin{equation}
  M_l^m
  =
  \sum\limits_{\lambda=0}^{l}
  \sum\limits_{\mu=-l}^l
  \tilde M_\lambda^\mu R_{l-\lambda}^{m-\mu}(\v{R}_0)
  =
  \left[
  \t{\tilde M}\circ\t{R}(\v{R}_0)
  \right]_l^m,
  \lbl{m2m}
\end{equation}
where $\tilde M_l^m=0$ if $|m|>l$. Eq.~\eref{m2m} defines the convolution operator $\circ$.

Direct evaluation of the telescoped sum of repeating cells requires to compute
the joint multipole moments. Let $\t{M}_\alpha$ be the multipole moment at stage
$\alpha$ of that telescoping process (with $\t{M}_0$ being the bare multipole
moment of the simulation cell). The multipole moment at stage $\alpha$ is then
given by the recursion formula
\begin{equation}
  \t{M}_\alpha
  =
  \sum\limits_\v{n}
  \t{M}_{\alpha-1}
  \circ
  \hat D(\mathcal{S}^{\v{N}_\alpha \v{n}}) \t{R}(\v{0}),
  \label{local_to_local}
\end{equation}
where $\v{n} = (n_1, n_2, \ldots)$ is the usual symmetry index and the sum runs from $-(N_i-1)/2$ to $(N_i-1)/2$ for each $n_i$. Furthermore, $\v{N}_\alpha \v{n}$ denotes
a component-wise multiplication and $\v{N}_\alpha = ( (N_1)^\alpha, (N_2)^\alpha,
  \ldots )$, hence summing up the contributions of $N_i$ repeat units for the symmetry $\mathcal{S}_i$. Note that $\v{N}_\alpha$ gives the distance of neighboring repeat units at stage $\alpha$ (in numbers of simulation cells).

Since $\hat D$ is an affine transformation we can split
it according to $\hat D = \hat T + \hat D_0$ [see also
  Eq.~\eref{coord_transf}], where $\hat T$ contains the translation operation
and $\hat D_0$ the linear transformation. We now absorb the translational contribution into the regular solid harmonics and rotate the multipole expansion. Eq.~(\ref{local_to_local}) then becomes
\begin{equation}
  \t{M}_\alpha
  =
  \sum\limits_\v{n} \hat D_0(\mathcal{S}^{-\v{N}_\alpha \v{n}})\t{M}_{\alpha-1}
  \circ
  \t{R}(\mathcal{S}^{-\v{N}_\alpha \v{n}} \v{0}),
  \label{local_to_local2}
\end{equation}
where $\mathcal{S}^{-\v{N}_\alpha\v{n}}\v{0}$ is the Cartesian distance between individual repeat units at stage $\alpha$. Intuitively, the multipole moment of a single stage in the telescoped expansion has to be given by a combination of rotated moments, which leads directly to Eq.~(\ref{local_to_local2}).

\emph{Multipole-to-local.} Let the multipole expansion $\t{M}$ be localized at
     $\v{0}$. Then the expansion $\t{L}$ of the local potential at $\v{R}_0$ is given by
\begin{equation}
  L_l^m
  =
  \sum\limits_{\lambda=0}^{l_\text{max}}
  \sum\limits_{\mu=-\lambda}^\lambda
  M_\lambda^\mu I_{l+\lambda}^{m+\mu}(\v{R}_0)
  =
  \left[
    \t{M} * \t{I}(\v{R}_0)
  \right]_l^m,
\end{equation}
where $*$ is another convolution operator.
In analogy to Eq.~(\ref{local_to_local2}) the local expansion of the potential at the center of the unit cell is given by
\begin{equation}
  \t{L}_l^m
  =
  \sum\limits_{\alpha=0}^{\alpha_\text{max}-2}
  \sum\limits_\v{n}
  \hat D_0(\mathcal{S}^{-\v{N}_\alpha\v{n}})
  \t{M}_\alpha
  *
  \t{I}(S^{-\v{N}_\alpha\v{n}}\v{0}),
\end{equation}
where the sum over $\v{n}$ now runs from $-N_i$ to $N_i$ excluding the terms
where the well-separateness criterion does not hold, \ie where $|n_i| \leq
\frac{N_i-1}{2}$ for any $i$. In total, this sums up $N_i^{\alpha_\text{max}}$ images of the simulation cell
for symmetry $\mathcal{S}_i$.

\emph{Local-to-local.} Let the local expansion $\t{\tilde L}$ be located at
      $\v{0}$. Then the expansion $\t{L}$ located at
     $\v{R}_0$ is given by
\begin{equation}
  L_l^m
  =
  \sum\limits_{\lambda=0}^{l_\text{max}}
  \sum\limits_{\mu=-\lambda}^{\lambda}
  \tilde L_{l+\lambda}^{m+\mu} R_\lambda^\mu(\v{R}_0).
\end{equation}
Note that since the local expansion is only applied within the simulation cell, no symmetry operation needs to be applied here.

\emph{Near-field.} The contribution of the neighboring $N_i$ repeat units to the electrostatic potential is computed by direct summation, as discussed in detail in Sec.~\ref{S:class_pot}.

\subsection{Transformation of multipole moments}

We apply a general linear transformation $\mathcal{L}$ to all coordinates. In particular, in $\mathcal{R}^3$ such linear transformation has a representation in a $3\times 3$ matrix $\t{T}$ and hence the transformed positions $\v{r}_i^\mathcal{L}$ are given by $\v{r}_i^\mathcal{L} = \mathcal{L} \v{r}_i = \t{T} \, \v{r}_i$. Under this transformation, the solid harmonics transform as $\hat D(\mathcal{L}) R_l^m(\v{r}_i) = R_l^m(\t{T} \, \v{r}_i)$ and from group-theoretical arguments \cite{Tinkham} it becomes clear that $\mathcal{L}$ has a block diagonal matrix representation in the function space of the solid harmonics. In particular,
\begin{equation}
  \hat D(\mathcal{L}) R_l^m(\v{r}_i) = \sum\limits_{m'=-l}^l \tilde
    D_{mm'}^{(l)} R_l^{m'}(\v{r}_i)
  = \tilde{\t{D}}^{(l)}(\t{T}) \, \v{R}_l(\v{r}_i)
\end{equation}
for each $l$ and $m$ with $|m|\leq l$. (Compare also Eq.~\eref{HnM} for
  the rotation of Hamiltonian and overlap matrix elements.) The computation of matrix $\tilde{\t{D}}^{(l)}$
up to arbitrary order in $l$ is described in Appendix~\ref{S:electrostatictr}.

\section{Application example: Polyalanine}

We are now in a position to use the RPBC in combination with the
self-consistent charge tight-binding model, using polyalanine in vacuum as an example system. This chiral
molecule belongs to the class of polypeptides, where the peptide unit
alanine has a single methyl sidegroup.
The RPBCs allow to continuously search for conformations of this molecule
without the need to choose conformations that are commensurate with a Bravais
unit cell~\cite{Ireta:2009p6069}.

The molecule is modeled using chiral symmetry.
We use the O-C-N-H tight-binding
parametrization of Elstner \emph{et al.}~\cite{ElstnerSCCDFTB}, augmented by
van-der-Waals
interactions~\cite{elstner_JCP_01} using the polarization data of
Miller~\cite{Miller1990}, and sample $\kappa$-space using
$20$ equally spaced $\kappa$-points.
The electrostatics interaction is computed using $l_\text{max}=8$, $N_1=11$ and $\alpha_\text{max}=5$.
For each conformation, we vary the recurrence length along the molecular axis
$z$ and optimize the chiral angle $\chi$.

\begin{figure}
  \center
  \includegraphics[width=9cm]{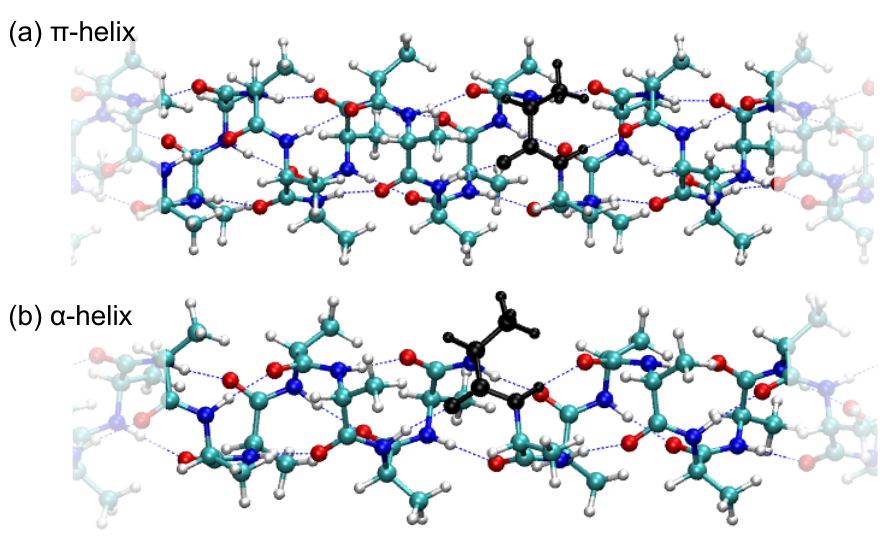}
  \caption{(Color online) Conformations of polyalanine in vacuum. (a) $\pi$-helix and (b)
    $\alpha$-helix. Atoms colored in black show the
    simulation cell consisting of $10$ atoms. Blue dotted lines indicate
    hydrogen bonds between the oxygen and hydrogen atoms.}
  \label{F:polyalanine}
\end{figure}

The equilibrium conformations
obtained for the $\pi$-helix and the $\alpha$-helix are
shown in Fig.~\ref{F:polyalanine}, where the atoms shown in black highlight
the atoms within a single unit cell. We find an equilibrium repeat length along the molecule's
axis of $1.21\,\text{\AA}$ and $1.57\,\text{\AA}$ for the $\pi$- and $\alpha$-helix, respectively, which
corresponds to a chiral angle of $82.5^\circ$ and $100.84^\circ$. This
compares well to full density-functional
calculations where a repeat length of
$1.17\,\text{\AA}$ and $1.50\,\text{\AA}$ is found for the $\pi$- and $\alpha$-helix, respectively~\cite{Ireta:2009p6069}.

Note that using a Bravais unit cell the chiral angle is given by
$\theta=360^\circ m/N$, where $m$ is the number of chain turns and
$N$ the number of residues per unit cell.
Hence, for $m=1$ the peptide's chiral angle is fixed to either
$90^\circ$ or $120^\circ$.
If the goal is to bracket the chiral angle of the $\alpha$-helix in an
interval smaller than $1^\circ$ the Bravais cell needs to contain at
least $m=32$ turns and $N=114$ residues, compared to a single residue
when using the RPBC approach.

\section{Conclusions}

We have demonstrated that RPBC is a simple but powerful atomistic simulation technique with easy implementation. It brings considerable efficiency gains and creates an access to distorted
material simulations. Implementations of RPBC in
quantum and classical codes will open possibilities
for new types of simulations, such as studies on
crack propagation under complex load conditions, DNA stretching, catalysis on curved surfaces,
and studies of macromolecules in nanochemical and nanobiological systems. We acknowledge that an implementation of RPBC in plane-wave DFT codes
is not straightforward and requires further theoretical research.
Nevertheless, the present paper shows that an implementation in
self-consistent charge NOTB is possible and it contains all the ingredients for an
implementation into full DFT codes that employ local basis sets and the fast
multipole method, such as for example Q-Chem.\cite{Shao:2006p3172}
We believe, that RPBCs, both in classical and quantum-mechanical formulation,
has the potential to become a truly useful
tool in computational nanoscience.

\vspace{2ex}
We would like to acknowledge Hannu H\"akkinen for support, Robert van Leeuwen for assistance, Aili Asikainen for graphics, Gianpietro Moras for an introduction into biomolecules, and  the Academy of Finland and the National Graduate School in Material Physics for funding. L.P. additionally thanks the University of Jyv\"askyl\"a for hospitality during a visit that was funded by the European Commission within the HPC-Europa2 project (grant agreement no 228398).

\appendix

\section{The transformation matrix %
$\mathbf{D}^{0}(\Rot{\aln})$}
\label{S:trmatrix}

In quantum mechanics, transformation of \wfs (\eg orbitals) upon rotation is found in group representation theory.
For atom's basis orbitals ordered as
\begin{align}
	 \ChemOhf,
	 \lbl{orbitals}
\end{align}
rotation representation has block-diagonal form
\def\DRotMMmatrixStructure{\begin{array}{c|c|c|c}
		D_0^{(0)} & 0              & 0             & \vTab\dots  \\\hline
		0          & \DRotMX{1}     & 0             & \vTab\dots  \\\hline
		0          & 0              & \DRotMX{2}    & \vTab\dots  \\\hline
		\vdots     & \vdots         & \vdots        & \ddots
\end{array}}
\begin{align}
	\DRotMM{\aln} = \left(\DRotMMmatrixStructure\right),
\end{align}
where
$\DRotMX{l}$
are $(2l+1)\times(2l+1)$ matrix blocks
that rotate orbitals with angular momenta $l= 0, 1, 2, \dots$.
Since the orientation of $s$-orbitals does not change upon rotation,
$D_0^{(0)} = 1$.
Since tight-binding basis orbitals are chosen real (see Ref.~\onlinecite{DFTB_Pekka_Ville}),
three $p$-orbitals, $p_x, p_y, p_z$, transform as $x, y, z$ coordinates:
$\DRotMX{1} = \Rot{\aln}$.

One way to obtain general expression for $\DRotMM{\aln}$ in tight-binding basis~\eref{orbitals} is to carry out similarity transformation of
the rotation representation, which
in commonly used basis of angular momentum eigenstates $\Ket{l, m}$ is given by~\cite{Rose, weissbluth}
\be
  D^{(l)}_{m'm} (\alpha, \Be, \gamma) = \exp{-\I m'\alpha}
  d^{(l)}_{m'm}(\Be)\exp{-\I m\gamma},
  \label{Wigd}
\ee
where $d^{(l)}_{m'm}$ are Wigner $\t{d}$-matrices
and $\alpha, \Be, \gamma$ are Euler's angles.
For $d$-orbitals, explicit expression for $\DRotMX{2}$ becomes tedious; we omit it here.

For illustration, however, we present a special case relevant for wedge and chiral setups.
For rotations about $z$-axis,
$p$-orbitals transform with
\def\DRotMMmatrixP{\begin{array}{ccc}
		       	\Caln  & -\Saln & 0     \\
				\Saln  & \Caln  & 0     \\
		       	0      & 0      & 1     \\
\end{array}}
\def\DRotMMmatrixD{\begin{array}{ccccc}
\CTaln & -\STaln & 0      & 0      & 0    \\
\STaln & \CTaln & 0      & 0      & 0     \\
0      & 0      & \Caln  & \Saln  & 0     \\
0      & 0      & -\Saln & \Caln  & 0     \\
0      & 0      & 0      & 0      & 1     \\
\end{array}}
\begin{align}
	\DRotMX{1}(\Rot{\aln}) = \left(\DRotMMmatrixP\right),	
        \label{porbtr}
\end{align}
and $d$-orbitals transform with
\begin{widetext}
\begin{align}
	\DRotMX{2}(\Rot{\aln}) = \left(\DRotMMmatrixD\right); 
\end{align}
\end{widetext}
note that it is the transpose of these matrices that is acting in Eq.~\eref{phi_n}.

\section{The transformation matrix $\t{\tilde{D}}^{(l)}(\t{T})$}\label{S:electrostatictr}

The matrices $\tilde{\t{D}}^{(l)}$ are related to the Wigner $\t{d}$-matrix, which describes the rotation of spherical harmonics, with normalization factors:
\begin{equation}
  \tilde D^{(l)}_{m m'} = \sqrt\frac{(l-m')! (l+m')!}{(l-m)! (l+m)!}
  d^{(l)}_{m m'};
  \label{Wigd2}
\end{equation}
note the difference between $\tilde{\t{D}}^{(l)}$ and $\t{D}^{(l)}$,  $\t{D}^{(l)}_0$ from previous section.

The $\tilde{\t{D}}^{(l)}$ matrix for $l=0$ and $l=1$ can be evaluated explicitly. In particular
$\tilde D^{(0)}_{0,0}=1$ and
\begin{widetext}
\begin{equation}
  \tilde{\t{D}}^{(1)}
  =
  \frac{1}{2}
  \begin{pmatrix}
    T_{yy} + T_{xx} + i ( T_{xy} - T_{yx} )
    &
    T_{xz} - i T_{yz}
    &
    T_{yy} - T_{xx} + i ( T_{xy} + T_{yx} )
    \\
    2(T_{zx} + i T_{zy})
    &
    2T_{zz}
    &
    2(- T_{zx} + i T_{zy})
    \\
    - T_{xx} + T_{yy} - i ( T_{xy} + T_{yx} )
    &
    - T_{xz} - i T_{yz}
    &
    T_{xx} + T_{yy} - i ( T_{xy} - T_{yx} )
  \end{pmatrix},
  \label{D1}
\end{equation}
where the latter one is the equivalent of Eq.~\eqref{porbtr} for the transformation of $p$-orbitals around $z$-axis.

In order to evaluate the matrices $\tilde{\t{D}}^{(l)}$ for $l\geq 2$ up to arbitrary order in $l$, we use a recursion scheme that has been developed for the rotation of spherical harmonics \cite{Choi:1999p8826,Ivanic:1996p6342}. In terms of the regular solid harmonics given by Eq.~(\ref{solid_harmonics_R}) the recursion becomes
\begin{align}
  \tilde D^{(l)}_{-l,n}
  =&
  \frac{1}{a_{l-1}^{-1,-l+1}}
  \left(
  a_{l-1}^{-1,n+1} \tilde D_{-1,-1}^{(1)} \tilde D^{(l-1)}_{-l+1,n+1}
  +
  a_{l-1}^{0,n} \tilde D_{-1,0}^{(1)} \tilde D^{(l-1)}_{-l+1,n}
  +
  a_{l-1}^{1,n-1} \tilde D_{-1,1}^{(1)} \tilde D^{(l-1)}_{-l+1,n-1}
  \right)
  \label{rec1}
  \\
  -l < m < l \;\text{:}\;
  \tilde D^{(l)}_{mn}
  =&
  \frac{1}{a_{l-1}^{0,m}}
  \left(
  a_{l-1}^{-1,n+1} \tilde D_{0,-1}^{(1)} \tilde D_{m,n+1}^{(l-1)}
  +
  a_{l-1}^{0,n} \tilde D_{0,0}^{(1)} \tilde D_{m,n}^{(l-1)}
  +
  a_{l-1}^{1,n-1} \tilde D_{0,1}^{(1)} \tilde D_{m,n-1}^{(l-1)}
  \right)
  \label{rec2}
  \\
  \tilde D^{(l)}_{ln}
  =&
  \frac{1}{a_{l-1}^{1,l-1}}
  \left(
  a_{l-1}^{-1,n+1} \tilde D_{1,-1}^{(1)} \tilde D^{(l-1)}_{l-1,n+1}
  +
  a_{l-1}^{0,n} \tilde D_{1,0}^{(1)} \tilde D^{(l-1)}_{l-1,n}
  +
  a_{l-1}^{1,n-1} \tilde D_{1,1}^{(1)} \tilde D^{(l-1)}_{l-1,n-1}
  \right),
  \label{rec3}
\end{align}
\end{widetext}
where, as opposed to a rotation of spherical harmonics~\cite{Choi:1999p8826,Ivanic:1996p6342}, Eqs.~(\ref{rec1})--(\ref{rec3}) hold for general linear transformations $\mathcal{L}$ with $\det\t{T} \not=1$ (\ie inversion or shear). The prefactors $a_l^{\alpha,m}$ are given by
\begin{align}
  a_l^{-1,m}
  =&
  \frac{(l-m+2)(l-m+1)}{2(2l+1)}
  \\
  a_l^{0,m}
  =&
  \frac{(l+m+1)(l-m+1)}{2l+1}
  \\
  a_l^{1,m}
  =&
  \frac{(l+m+2)(l+m+1)}{2(2l+1)}.
  \label{ap1}
\end{align}
Equations~(\ref{D1}) to (\ref{ap1}) are the explicit representation of the rotation operation.



\end{document}